\documentclass[final,5p,times]{elsarticle}

\usepackage[utf8]{inputenc}

\usepackage{graphicx}
\usepackage{epsfig}

\usepackage{bm,amssymb,amsmath,amsfonts,xcolor}

\biboptions{sort&compress}

\journal{Biochimica et Biophysica Acta -- Biomembranes}

\begin{document}

\begin{frontmatter}



\title{pH Sensing by Lipids in Membranes: The Fundamentals of pH-driven Migration, Polarization and Deformations of Lipid Bilayer Assemblies}


\author[label1,label2]{Miglena I. Angelova} 
\author[label3]{Anne-Florence Bitbol} 
\author[label2]{Michel Seigneuret} 
\author[label4]{Galya Staneva} 
\author[label5]{Atsuji Kodama} 
\author[label5]{Yuka Sakuma}
\author[label5]{Toshihiro Kawakatsu} 
\author[label5]{Masayuki Imai}  
\author[label1,label2]{Nicolas Puff} 

\address[label1]{Sorbonne University, Faculty of Science and Engineering, UFR 925 Physics, Paris F-75005, France}
\address[label2]{University Paris Diderot - Paris 7, Sorbonne Paris Cité, Laboratory Matière et Systèmes Complexes (MSC) UMR 7057 CNRS, Paris F-75013, France}
\address[label3]{Sorbonne University, Faculty of Science and Engineering, Laboratory Jean Perrin, UMR 8237 CNRS, Paris F-75005, France}
\address[label4]{Institute of Biophysics and Biomedical Engineering, Bulgarian Academy of Sciences, Sofia, Bulgaria}
\address[label5]{Department of Physics, Tohoku University, Aoba, Sendai 980-8578, Japan}

\begin{abstract}
Most biological molecules contain acido-basic groups that modulate their structure and interactions. A consequence is that pH gradients, local heterogeneities and dynamic variations are used by cells and organisms to drive or regulate specific biological functions including energetic metabolism, vesicular traffic, migration and spatial patterning of tissues in development. While the direct or regulatory role of pH in protein function is well documented, the role of hydrogen and hydroxyl ions in modulating the properties of lipid assemblies such as bilayer membranes is only beginning to be understood. Here, we review approaches using artificial lipid vesicles that have been instrumental in providing an understanding of the influence of pH gradients and local variations on membrane vectorial motional processes: migration, membrane curvature effects promoting global or local deformations, crowding generation by segregative polarization processes. In the case of pH induced local deformations, an extensive theoretical framework is given and an application to a specific biological issue, namely the structure and stability of mitochondrial cristae, is described.
\end{abstract}


\end{frontmatter}



\section{Introduction}
Throughout their life, cells are submitted to an inhomogeneous and variable environment, to which they adapt and respond. In particular, local pH inhomogeneities at the cellular scale are ubiquitous and very important, e.g. in migrating cells~\cite{Martin11} and in mitochondria~\cite{Davies11}. While cellular response to complex molecules are mediated by dedicated biochemical signaling pathways, e.g. in chemotaxis~\cite{Sourjik12}, cellular response to pH may also involve more universal physico-chemical mechanisms. Indeed, membrane lipids are directly affected by pH, due to their acido-basic properties. Such chemical modifications of lipids have generic physical effects on the cell membrane.

In order to quantitatively investigate these generic effects of pH on lipid bilayer membranes, experiments on biomimetic membranes~\cite{Angelova99} are particularly useful. A line of our work has focused on studying the various effects of pH changes on giant unilamellar vesicle (GUV). We demonstrated that a pH change can induce vesicle migration and global deformation~\cite{Kodama2016}, as well as polarization in vesicles involving phase-separated membrane domains~\cite{Staneva2012}. Furthermore, we showed that localized pH heterogeneities can induce local dynamical membrane deformations and developed a theoretical description of this phenomenon~\cite{Khalifat08,Fournier09,Khalifat11,Bitbol12_guv}. We were also able to mimic the formation and behavior of healthy and diseased mitochondrial cristae using purely lipidic biomimetic membranes~\cite{Khalifat08,Khalifat12,Khalifat14}. In this paper, we review this work, and we discuss its context and implications.

\section{pH-induced migration, deformation and polarization processes in cells}
The ability of cells to detect and react to chemical concentration gradients is instrumental for cell motility, proliferation and differentiation, and therefore essential for processes such as immunological response, development and wound healing. The mechanisms underlying cell chemotaxis toward complex organic or biosynthetic molecules have been widely documented~\cite{eisenbach2004chemotaxis}. Additionally, the role of gradients of ``morphogens", i.e. secreted signaling molecules that determine the arrangement and developmental fate of cells according to their position inside a developing tissue or organism, is now recognized~\cite{Inomata2017}. The role of pH gradients, which involve the ubiquitous H$^+$ (H$_3$O$^+$) and OH$^-$ ions, in cell motility and deformation processes, emerged more recently. Proton gradients across membranes are well characterized as an essential intermediate electrochemical potential form for ATPase/synthase-associated bioenergetic processes in mitochondria, chloroplasts and unicellular organisms~\cite{Mitchell2011}. Also well-known is the role of intracellular vesicle transmembrane pH gradients as regulatory factors in endosomal, lysosomal and secretory pathways~\cite{Hu2015}. However, more recently, it has been found that pH gradients also act as chemotactic cues for microorganisms such as bacteria or amoeba or even sperm and osteoblastic cells~\cite{Yang2012, Korohoda1997, Iida2017, Kirchhof2011}. Other specific roles of local pH gradients and heterogeneities, either intracellular, extracellular or transmembrane are currently being discovered. 

For eukaryotic cells, both external and cytoplasmic pH are globally uniform and near-neutral. Maintaining cytoplasmic pH is essential for optimal cellular metabolism, growth and proliferation~\cite{Casey2009}. However, there are specific situations where such pH uniformity may be globally or locally affected. For instance, dynamic changes of the cell interior pH, within a range of 0.2 to 0.3 units, function as a signaling mechanism to regulate a number of cellular processes such as cell cycle progression, proliferation/apoptosis and differentiation~\cite{Srivastava2007}. Spatial patterning of pH over different cell types occurs during oogenesis and is supposed to play important roles in development by influencing cell division, modifying the cytoskeletal organization and regulating migration processes~\cite{Krueger2015}. Both internal and external pH abnormalities are hallmarks of cancer cells and of tumor progression~\cite{White2017}. A highly polarized cell tissue for which a spectacular transverse pH gradient (2.5 units over 10 $\mu$m) occurs is the stratum corneum~\cite{Hanson2002} i.e., the upper layer of the epidermis. Besides, even under conditions where bulk pHs are stable, internal and external pH can also be locally and dynamically affected, particularly in juxtamembrane regions~\cite{Casey2009}. Such situations are often related to a nonuniform distribution or activation of the Na$^+$/H$^+$ exchangers (NHE) in the plasma membrane~\cite{Grinstein1993, Klein2000}. In animal cells, cytosolic pH is primarily regulated by these exchangers, which export protons from the cytoplasm through facilitated exchange with extracellular Na$^+$~\cite{Casey2009}. Along with such bulk cytoplasmic pH standard regulation, localized NHE activities in regions of the plasma membrane also generate pH microdomains that regulate the actin cytoskeleton in cell polarization events~\cite{Li2014, Ro2004, Frantz2007}. Increased NHE activity in the growth cone of neurons compared with the cell body causes elevated local cytoplasmic pH, which results in increased polarization/extension of neurites~\cite{Sin2009}. The NHE is also upregulated in post-synaptic membranes during neuronal activity where it negatively regulates dendritic spine growth~\cite{Diering2011}. NHE-dependent local pH heterogeneities or lateral gradients have also been directly measured in oligodendrocytes~\cite{Ro2004}. Migrating cancer cells and polarized epithelial cells also display NHE-dependent external pH gradients between the apical, lateral and basal sides~\cite{Stuewe2007, Stock2007, Martin_2010, GondaG259, MaouyoC973, Campetelli2012}. Transcellular pH gradients exist in large polarized cells such as pollen tubes or fucoid eggs~\cite{Gibbon1994, Feijo1999}. These pH gradients are in fact likely to be cortical or submembranous too, since the rapid diffusion of protons may impair the formation of bulk cytoplasmic gradients.

A role of such local dynamic pH gradients is exemplified in those cell migration, global deformation and asymmetric growth processes which are associated with the polarization of a leading edge along a large region of the plasma membrane. Aside from the role of cytoskeletal components, small GTPases and membrane microdomains, recent studies have identified the importance of certain proton transporters in regulating such processes. Local pH gradients at the plasma membrane appear to be necessary for actin filament assembly and focal adhesion remodeling~\cite{Campetelli2012, Schoenichen2013, Srivastava2008, Choi2013, Frantz2008, Haupt2014}. In specific polarized migrating cells, the Na$^+$/H$^+$ exchanger 1 (NHE1) has been shown to be necessary for migration~\cite{Denker2002, Putney2002} , and its segregation at the lamellipodium generates both extracellular and intracellular surface H$^+$ gradients from the rear end to the leading edge~\cite{Stuewe2007, Stock2007, Martin_2010}. 

Interestingly, some of the functional processes associated with such pH heterogeneities have been linked to protonation changes of acidic phospholipids, affecting either interactions with signaling proteins~\cite{Simons2009} or membrane curvature~\cite{Ben-Dov2012}. The lipid phosphatidic acid (PA) has important roles in cell signaling and me\-tabolic regulation in all organisms. New evidence indicates that PA also has a crucial role as a pH biosensor, coupling changes in pH to intracellular signaling pathways~\cite{Shin2011}. For instance, local pH regulation and phospholipid charges have been implicated in planar cell polarity pathway regulation in Drosophila epithelial tissues~\cite{Simons2009}. This polarity relies on the targeting of the Dishevelled (Dsh) protein complex to the plasma membrane for binding to the transmembrane protein Frizzled (Fz) at cell-cell contacts. Local juxtamembrane  pH values are regulated by a putative Na$^+$/H$^+$ exchanger NHE2, affecting acidic phospholipid (presumably phosphatidic acid) headgroup protonation levels and impacting the binding of the polybasic stretch of Dsh to the plasma membrane inner leaflet. Slightly alkaline local pH conditions thereby trigger the ``non-canonical" Fz/PCP pathway. These studies thus demonstrate the importance of proton transporters in local pH regulation, and of membrane lipid charges in the proper stabilization of a polarity axis in a tissue context.

Other likely candidates as local pH biosensors are the signaling phospholipids phosphoinositides (PIPs). Signaling proteins bind to PIPs generally through structurally conserved binding domains~\cite{Stahelin2008}. Migrating cells establish a polarization axis by specifying a zone of actin polymerization at the leading edge. This process is regulated by the polarized recruitment and activation of polarity factors such as the conserved Rho-type small GTPase cdc42. Local activation of cdc42 is mediated by guanine exchanging factors (GEFs) that are recruited through interactions with phosphatidylinositol 4,5-bisphosphate (PIP2). In fibroblasts, the so\-dium/proton exchanger NHE1 promotes polarity and migration. This transporter, through its regulatory effect on local submembrane pH, may promote binding of a cdc42 GEF to PIP2 at the leading edge plasma membrane, possibly by PIP2 protonation change.

Gangliosides constitute another type of negatively-charged lipids. More specifically, they are glycosphingolipids that bear a protonable sialic acid moiety. These are present in lipid rafts and may also form distinct specific microdomains~\cite{Hullin-Matsuda2007, Hakomori2000, Gupta2009}. Both rafts and ganglioside-specific microdomains are involved in numerous cellular activation, transduction, and signaling functions. Cellular processes associated with raft/gangliosides microdomains often involve their clustering and aggregation. Two main types of mechanisms have been considered and documented as responsible for such microdomain aggregation~\cite{Simons2010}: (i) cross-linking of microdomain membrane components by multivalent extracellular ligands and (ii) dragging of microdomains by cytoskeletal rearrangement through specific cytoskeleton-membrane interaction. However, an interplay of local juxtamembrane pH effects and ganglioside topography may also be involved. Incidentally, polarized segregation of raft and/or ganglioside microdomains does occur in migrating cells~\cite{Sitrin2010, Manes1999, Gomez-Mouton2001} and differences in lipid rafts and ganglioside distribution between the apical and basal membranes have also been established for epithelial cells. Besides, NHE1 partially colocalizes with GM1 ganglioside-containing rafts~\cite{Yi2008}.
 An effect of pH on the dynamics, size, and lateral distribution of membrane domains containing acidic lipids such as gangliosides has been observed in model membrane systems (see Section~\ref{pola} and~\cite{Staneva2012}). Recently, in LN229 glioblastoma cells, changes in extracellular pH have been shown to induce differential clustering of surface gangliosides in correlation with cell apoptosis. Such a process may occur in situ, since extracellular tumor acidification is a well-recognized hallmark of cancer progression~\cite{John2017}.   

A more direct effect of a pH gradient on cells, also linked to titration of membrane lipids, is ``proton-induced endocytosis"~\cite{Ben-Dov2012, Ben-Dov2013}. It has been found that receptor-independent endocytic-like events can be triggered by exposing cells to external pH $<$ 6. The suggested mechanism is protonation of negatively charged lipids in the plasma membrane outer leaflet, reducing their electrostatic repulsion and consequently the area occupied per molecule. The mechanical imbalance that thereby develops across the plasma membrane in terms of surface area can be balanced by a higher local curvature of the membrane. It has been suggested that  membrane invaginations develop at sites where line tension pre-exists, i.e. at  membrane microdomains. Changes in lipid packing were found to modulate the endocytosis, presumably by affecting such line tension.

\section{Effect of pH gradients on vesicle global motions: migration, global deformations and polarization}

Numerous studies have been conducted to investigate the effect of proton gradients on the motional properties of model membranes. Most studies have involved local or global external acidification of vesicles prepared by various methods. Processes that have been observed include vesicle migration, vesicle global and local shape changes and heterogeneous vesicle ``polarization" (i.e. vectorial segregative clustering of specific bilayer components). In this section, emphasis will be given to processes involving vesicles globally, i.e. migration, overall vesicle deformations, global polarization. pH effects promoting local shape changes will be considered in sections~\ref{local}  and~\ref{biorel}.

\subsection{pH-induced migration}
\label{migration}

\paragraph{Physical mechanisms involved in vesicle migration}
The mechanisms involved in solute gradient-induced migration of a particle (solid particle, droplet, vesicle) have been widely studied~\cite{Anderson1982, Prieve1984, Hanczyc2007, Tsemakh2004}. However, the relevance of such mechanisms for the migration of lipid vesicles are still a matter of debate. These processes might be grouped in two categories: (i) processes in which the solute gradient only creates an external force that acts on the particle and promotes its migration; (ii) processes which involve a chemical interaction of the solute with the particle (binding or chemical reaction).

Phenomenologically, the non-specific mechanisms correspond to processes known as osmophoresis and electrophoresis (also called diffusiophoresis). Osmophoresis arises from the fact that the solute concentration gradient underlies a hydrostatic pressure gradient. There is therefore an imbalance of hydrostatic pressure around the particle. The hydrostatic force propels the particle in the direction opposite to the gradient. If the particle is charged and the solute is an electrolyte, electrophoresis also takes place. The latter is due to the difference in the diffusion coefficients of the anion and the cation, that creates a local electric field on the particle. The direction of the electrostatic force that propels the particle depends on the mobilities and valences of anions and cations. In the case of lipid vesicles, this mechanism can be significant even if no charged headgroup is present. For example zwitterionic phosphatidylcholine vesicles can be propelled by large gradients of various metal chlorides due to the exterior surface dipole potential associated with the specific orientation of the phosphocholine headgroup~\cite{Kodama2017}. 

Specific mechanisms of solute gradient-induced particle migration primarily involve the surface tension of the particle. If the solute interacts or reacts chemically with the particle surface, a consequence of the gradient is an underlying spatial variation of the extent of the interaction or of the reaction along this surface. As a result, a gradient of surface tension may occur on the particle. Such gradient can promote migration of the vesicle by two possible distinct mechanisms. The simplest mechanism is just thermodynamic, i.e. the particle tends to diffuse down its chemical potential gradient. The asymmetric distribution of the surface tension on a vesicle results in a force, which drives the vesicle along the axis of the solute gradient with a direction corresponding to increased stability. For non-solid particles (droplets and, potentially, vesicles), a second mechanism may also occur. It is the so-called ``Marangoni convection". The particle reacts by moving in the direction of the solute gradient.

For lipid vesicles in a pH gradient, two types of particle/solute chemical interactions or reactions are possible: (i) titration of acidic or alkali headgroup functions (i.e. binding or dissociation of H$^+$ or OH$^-$); (ii) chemical reactions involving or catalyzed by H$^+$ or OH$^-$. The relevance of the two overmentioned surface tension mechanisms for lipid vesicle propulsion have been investigated. In particular, it has been claimed that a possible method to detect the presence of a Marangoni flow is to use lipid domain-containing vesicles that can be imaged by fluorescence microscopy. The flow, if existing, should occur inside the membrane and promote a visible vectorial motion of domains. Results on 2 types of vesicle migration phenomena were unable to detect such a motion on lipid vesicles which behaved as hard shells~\cite{Kodama2017, Kodama2016}. On the other hand, Marangoni effects have been found responsible for local deformations (see Section~\ref{local}).

\begin{figure}[htb]
	\begin{center}
		\includegraphics[width=1\columnwidth]{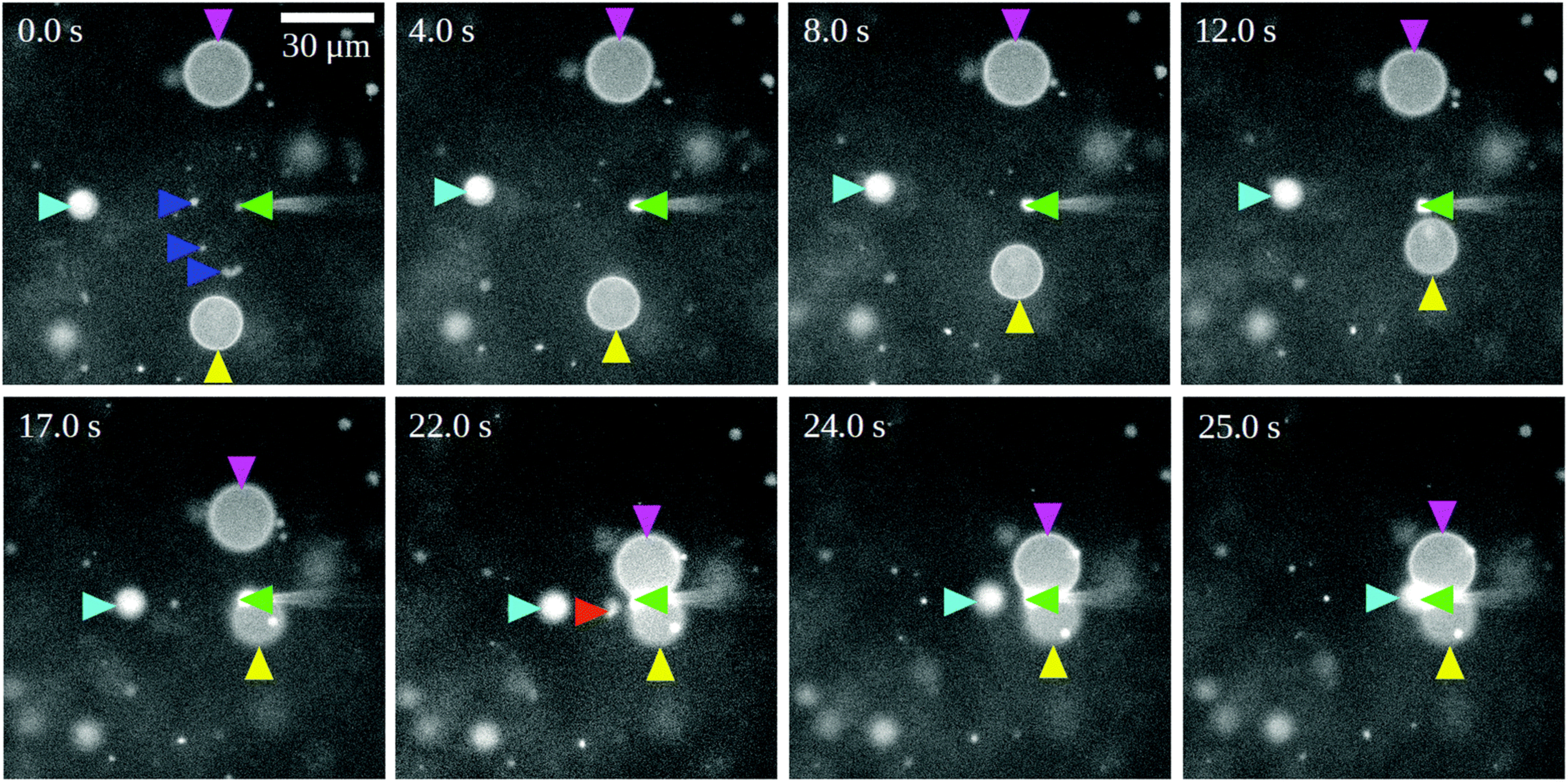}
		\caption{Snapshot images of migrating DOPC-GUVs labeled with Rh-DOPE triggered by the micro-injection of a 10 mM NaOH solution. GUVs are formed in pure water. The elapsed time since the start of the microinjection (t = 0.0 s) is shown at the top corner of each image. The representative GUVs are marked with magenta, cyan, and yellow arrowheads in each image, and the tip of the micropipette is also marked with a green arrowhead. The small lipid assemblies near the micro-pipette are marked with dark blue arrowheads (0.0 s frame), and a GUV from an out-of-focus plane is marked with an orange arrowhead (22.0 s frame). The scale bar in 0.0 s frame is 30 $\mu$m. \textit{Reprinted from Kodama et al.~\cite{Kodama2016} with permission of the Royal Society of Chemistry.}}
		\label{fig_migration}
		\end{center}      
\end{figure}

\paragraph{Experimental demonstration of pH gradient-induced vesicle migration}
\label{expmig}
Recently, we demonstrated the migration of giant phospholipid (DOPC, 3-10 $\mu$m) vesicles in response to a pH gradient~\cite{Kodama2016}. Upon microinjection of a NaOH solution (10 mM, pH 12), giant unilamellar vesicles (GUVs) linearly move toward the tip of the micropipette (Fig.~\ref{fig_migration}). The elapsed time before each vesicle starts to move is dependent upon its initial distance from the micropipette tip. When the injection is stopped, the motion of the vesicles stops within 1s. Under the same experimental conditions, no migration is observed with monovalent salt, acid or sucrose solutions, indicating that the observed migration of vesicles is only driven by the concentration gradient of the hydroxide ion. The electrolyte concentration dependence of the vesicle mobility was found to be inconsistent with diffusiophoresis theory. On the other hand, separate data indicated that NaOH promotes a time-dependent decrease of the surface tension of DOPC membranes. This suggested that the NaOH-induced vesicle mobility is due to the formation of a surface tension gradient along the vesicle, due to the pH gradient. Marangoni propulsion was excluded since similar migration experiments with vesicles containing L$_o$ domains did not show any unidirectional flow affecting domain motion. Two mechanisms were tentatively investigated for the alkaline pH-induced surface tension decrease: (i) titration of acidic or alkali headgroup functions; (ii) alkaline hydrolysis of PC to lyso-PC and fatty acid. The measured vesicle migration velocity vs. distance to micropipette tip dependence was found to be more in favor of the second mechanism where the hydrolysis of the phospholipids by NaOH decreases the surface tension of the vesicle. More generally, the migration of vesicles in a pH gradient is qualitatively described by a surface tension gradient model where the vesicles move toward the direction of lower surface energy. The gradient produces an effective force to propel the vesicles.

\subsection{pH-induced vesicle global deformation}
\label{deform}

\begin{figure}[htb]
	\begin{center}
		\includegraphics[width=1\columnwidth]{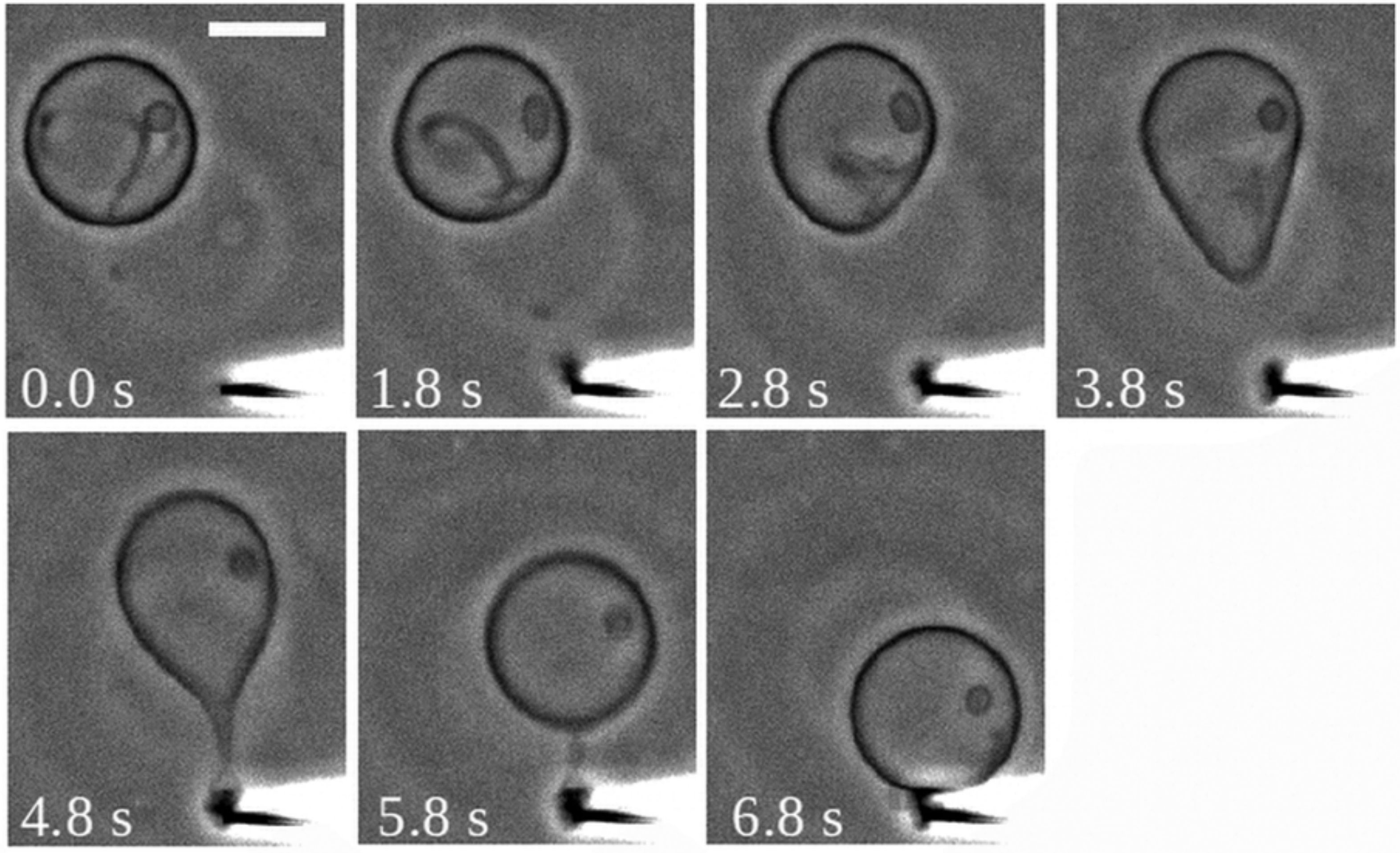}
		\caption{Simultaneous migration and global deformation of a single DOPC-GUV labeled with Rh-DOPE subjected to microinjection with a 10 mM NaOH solution. The GUV is formed in pure water. The elapsed time after the injection is shown. The scale bar indicates 20 $\mu$m. \textit{Reprinted from Kodama et al.~\cite{Kodama2016} with permission of the Royal Society of Chemistry.}}
		\label{global}
	\end{center}      
\end{figure}

Bulk or transmembrane pH gradients can promote global deformation of lipid vesicles by extremely diverse mechanisms. For instance, all the mechanisms that sustain migration of a vesicle can also promote its global deformation provided that the vesicle is ``deflated", i.e. has a subspherical volume to surface ratio. This is illustrated in Fig.~\ref{global}, which corresponds to the experiment described in~\ref{expmig}~\cite{Kodama2016}. Here the pulling force associated with the pH gradient-induced surface tension asymmetry causes not only the migration but also the shape deformation of the vesicle (the vesicle initially has an invagination indicating its flattened character). Immediately after the NaOH micro-injection, the vesicle begins to move, and simultaneously the membrane starts to deform to a teardrop shape toward the tip. This is due to the combined effects of the pulling force, vesicle deformability and solvent resistance. At later times, from the apex of the teardrop vesicle, a tubular membrane protrudes toward the tip and bridges between the tip and the vesicle. This is a local deformation, the mechanism of which will be dealt with in Section~\ref{local}.

Other mechanisms of pH-induced global deformation of li\-pid vesicles have been described as well. For instance, it has been observed that external acidification induces shape transformation of phosphatidylglycerol-containing vesicles due to transmembrane pH gradient-driven vectorial inward transverse diffusion of the later lipid~\cite{Farge1992} (this mechanism will be detailed in Section~\ref{biorel}). Of particular interest is the work of Shioi and collaborators who described oscillatory shape changes or amoeboid-like shape changes of double layered oleic acid/sodium oleate vesicles upon external alkalization~\cite{Nawa2015, Nawa2013}. The complex mechanisms involved various processes : ion permeation, transient pore formation, osmotic swelling and membrane elastic response.

\subsection{pH-induced vesicle chemical polarization}
\label{pola}

\begin{figure}[htb]
	\begin{center}
		\includegraphics[width=1\columnwidth]{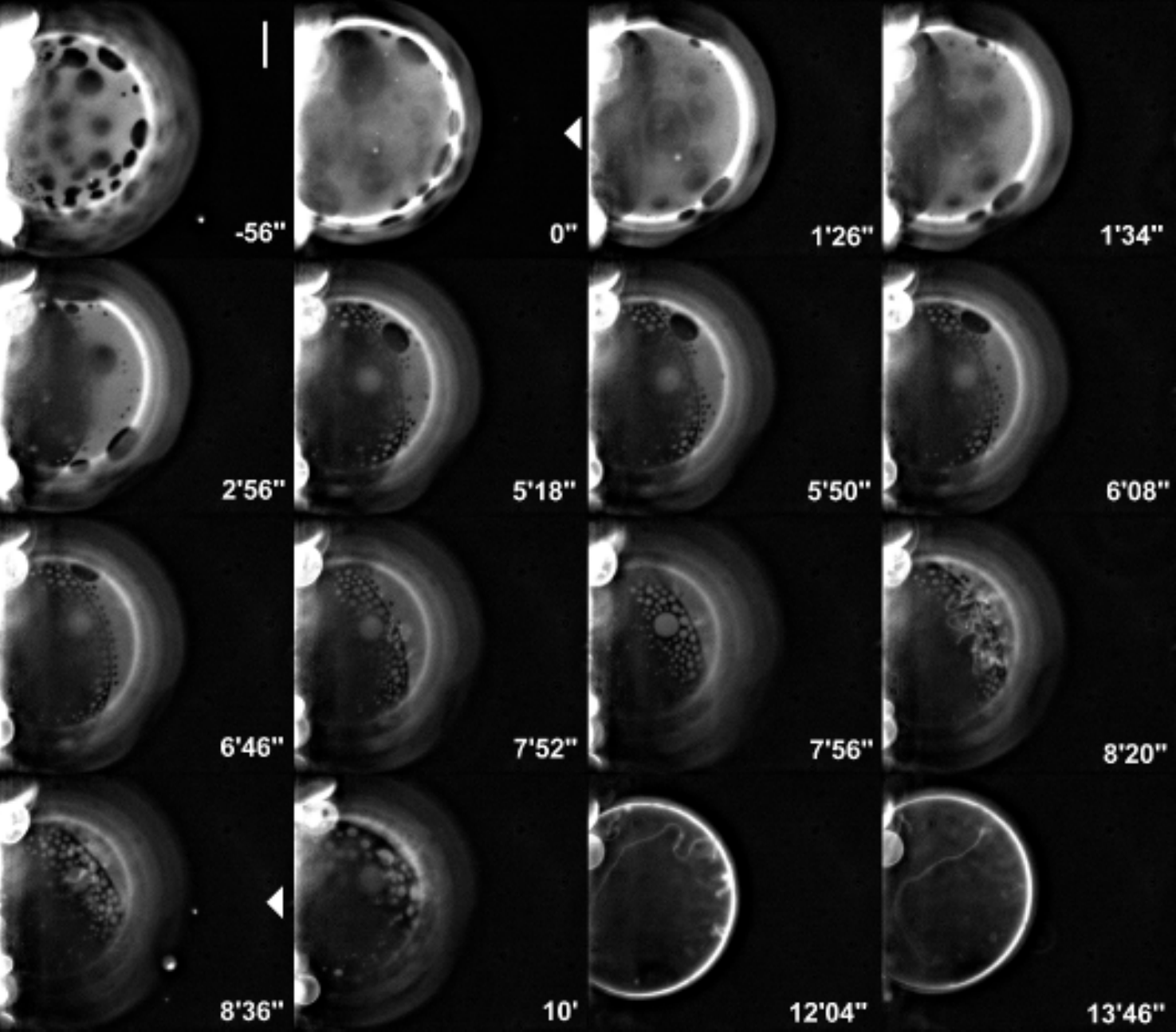}
		\caption{Fluorescence microscopy images showing the time-dependent development of lateral segregation, vesicle polarization, and L$_d$ phase tubulation on photoinduced L$_o$/L$_d$ phases after local external acidification by 100 mM HCl injection (25 hPa pressure) at 28$^{\circ}$C on a single GUV with molar composition EggPC/EggSM/CHOL/Ovine brain GM1 50:28:20:2 (mol/mol) labeled with TR-PE (0.25 mol\%), formed in a buffer (HEPES 0.5 mM, pH 7.4). The times before or after the beginning of acidification are indicated for each image as negative or positive values respectively. White arrows indicate the position of the micropipet tip as well as the start and end of acid microinjection. Bar: 20 $\mu$m. \textit{Reprinted from Staneva et al.~\cite{Staneva2012} with permission of ACS Publications.}}
		\label{fig_pola}
	\end{center}      
\end{figure}

The fact that specific lipids segregate into domains as a response to pH changes is well known. However, currently, only one study has addressed the possibility of inducing polarization, i.e. vectorial segregation of specific membrane components, in a vesicle by pH gradients~\cite{Staneva2012}. In order to induce such an effect, the pH gradient must be focused on one side of the vesicle. We have studied the influence of such a lateral pH gradient along the membrane surface on lipid microdomain dynamics in giant unilamellar vesicles containing phosphatidylcholine, sphingomyelin, cholesterol, and the ganglioside GM1. L$_o$/L$_d$ phase separation was generated by photosensitization~\cite{Staneva2011}. In this approach, the membrane-embedded marker Texas Red-PE is used both as a photosensitizer and a fluorescence probe imaging L$_o$ domains, from which it is excluded~\cite{Bagatolli2006}. Photosensitization that occurs due to the illumination light used for fluorescence observation of GUVs under the microscope, generates modified lipids that promote the formation of an L$_o$ phase. The latter is visible due to its exclusion of Texas Red-PE. Depending on temperature and composition, the process ultimately yields L$_o$ domains over a continuous L$_d$ phase (dark spots over white background) or the opposite. In order to establish a spatial lateral pH gradient at the GUV membrane surface, we performed local delivery of an acid solution (100 mM HCl) in the vicinity of the vesicle using a micropipette. Calculation of the steady-state pH at the membrane surface yielded values of 4.75 to 5.0. Microinjection was started at a specific stage of photoinduced L$_o$ phase separation corresponding to evenly distributed round-shaped microdomains of 5--10 $\mu$m size. At 28$^{\circ}$C, with a morphology initially consisting of discrete photoinduced micrometric L$_o$ domains over a continuous L$_d$ phase, there is a definite segregative movement of both phases in opposite directions with respect to the proton source. A clustering and coalescence of L$_o$ domains toward the region of lower H$^+$ concentration, opposite to the side of the vesicle where acid is injected, is evident, leaving a cap of nearly pure L$_d$ phase on the acid-injection side. This process ultimately gives rise to a directionally ``polarized" vesicle with an L$_o$ pole located toward the higher pH side together with an L$_d$ pole located toward the lower pH side, aligned in the axis of the proton source (Fig.~\ref{fig_pola}). Once such polarization of the vesicle into two continuous poles is established, a later process is the protrusion of inward tubes from the L$_d$ phase lipids toward the vesicle interior. When acid injection is switched off, the vesicle remains bipolar but the L$_d$ and L$_o$ poles move away from their directionally polarized position with regard to the micropipette. 

\begin{figure}[htb]
	\begin{center}
		\includegraphics[width=1\columnwidth]{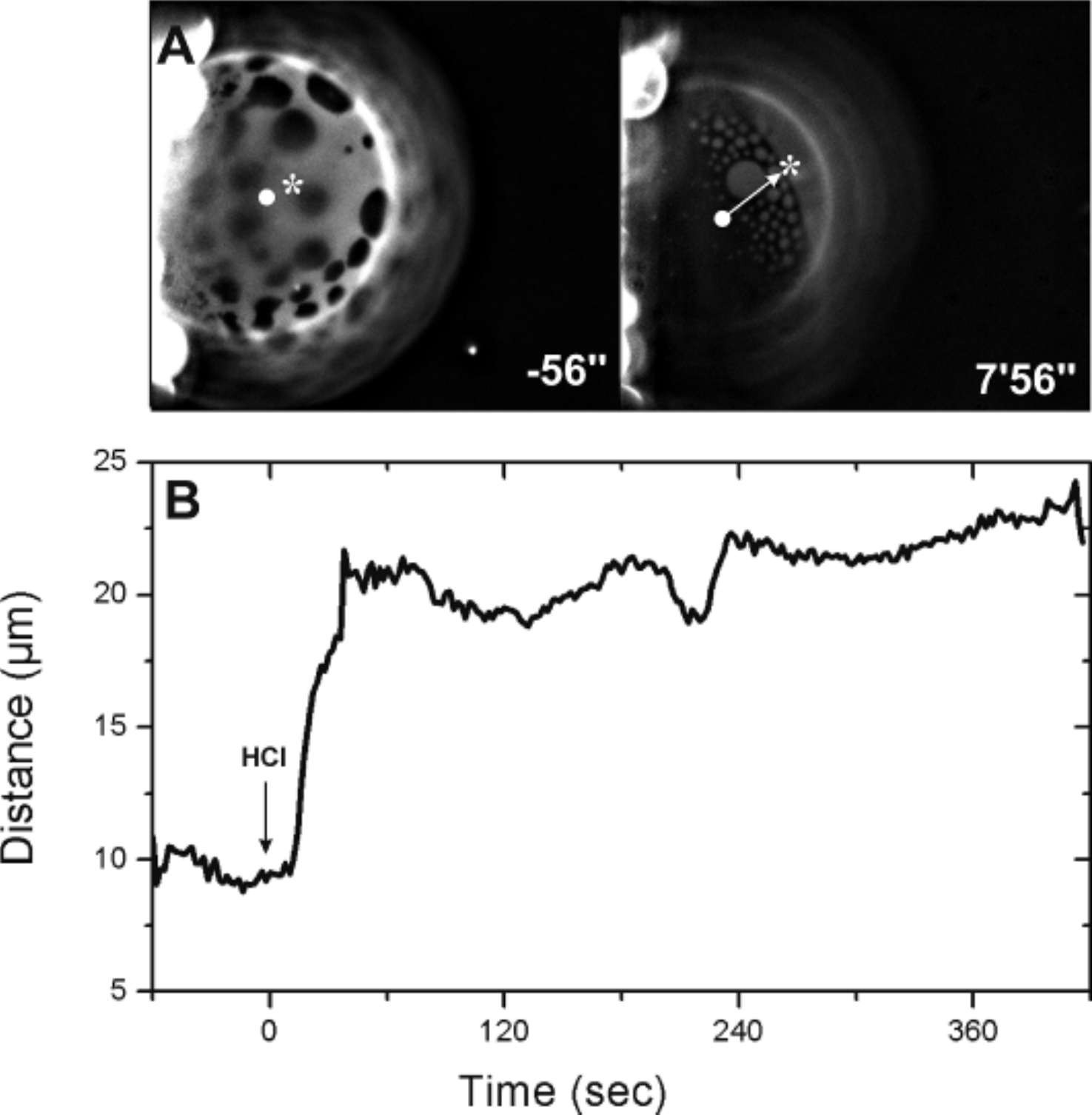}
		\caption{Quantification of lateral segregation and vesicle polarization on photoinduced L$_o$/L$_d$ phases after local external acidification at 28$^{\circ}$C on a single GUV with molar composition EggPC/EggSM/CHOL/Ovine brain GM1 50:28:20:2 (mol/mol) labeled with TR-PE (0.25 mol\%), formed in a buffer (HEPES 0.5 mM, pH 7.4), corresponding to the data of Figure~\ref{fig_pola}, from the evolution of the centroids (barycenters) of bright and dark pixels. A. Position of the centroids for dark (white circle) and bright (white star) ca. 1 min before and ca. 8 min after local external acidification. B. Time evolution of the distance between both centroids after acid injection. \textit{Reprinted from Staneva et al.~\cite{Staneva2012} with permission of ACS Publications.}}
		\label{bary}
	\end{center}      
\end{figure}

Figure~\ref{bary} shows the kinetics of lateral segregation and polarization of L$_o$/L$_d$ phases. There is a ca. 1 min rapid phase followed by a slower phase that corresponds respectively to the initial segregation of domains and their subsequent coalescence. Similar results were obtained at 20 $^{\circ}$C with an initial photoinduced phase separation morphology consisting of discrete L$_d$ domains over a continuous L$_o$ phase, i.e., inverse from that found at 30 $^{\circ}$C. None of these processes occurs without GM1 or with the uncharged analog asialo-GM1. These are therefore related to the acidic character of the GM1 headgroup. Control LAURDAN fluorescence experiments on large unilamellar vesicles (LUVs) indicated that, with GM1, an increase in lipid packing occurs with decreasing pH, attributable to the lowering of repulsion between GM1 molecules. Packing increase is much higher for L$_d$ phase vesicles than for L$_o$ phase vesicles. The local lateral pH gradient-induced segregation of L$_o$ and L$_d$ domains can be interpreted as follows. The lateral pH gradient imposed to the heterogeneous GUV may give rise to a spatial gradient of chemical potential for both types of domains/phases, which acts as an effective force for segregation. Low pH, which protonates GM1 molecules and abolishes their mutual repulsion, promotes a much higher condensing and ordering effect on the L$_d$ phase than on the L$_o$ phase. This may lead to a lower free energy of L$_d$ lipid domains at the acid side of the gradient (due to decrease of electrostatic repulsion and, possibly, stronger interactions between lipid molecules) and therefore to diffusion and gathering of L$_d$ domains/phase toward this region of lower free energy. The inward tubulation effect of local acidification can be interpreted as resulting from GM1 protonation yielding changes in both lipid local density and spontaneous curvature in the outer leaflet as described in detail in Section~\ref{biorel}.

\section{Deformation of a membrane in response to a local chemical modification}
\label{local}

In the previous section, we described different global responses of vesicles to pH heterogeneities, namely migration, global deformation and polarization. We are now going to discuss the physics of the local bilayer membrane deformations caused by pH heterogeneities.

\subsection{Context}
\label{context}

\paragraph{Static and global modifications}

The response of a vesicle to a static and uniform modification of its environment is well described by the area-difference elasticity (ADE) model~\cite{Svetina85,Miao94,Seifert_book}. This model goes beyond the classic Helfrich model based on membrane curvature energy~\cite{Helfrich73} by including an extra term quadratic in the difference of area between the two monolayers. Qualitatively, each monolayer potentially has a different preferred area, determined by chemical composition and number of lipids. However, for these preferred areas to be adopted simultaneously, the membrane may need to curve, which costs bending energy. The actual equilibrium vesicle shape is thus determined by a compromise between bending energy and area-difference elasticity energy.

Within the ADE model, the equilibrium shape of a vesicle is determined by two dimensionless parameters~\cite{Miao94}. The first one is its reduced volume $v$, defined from the volume and the area of the vesicle. The second one, $\overline{\Delta a_0}$, is a sum of a contribution from the preferred area difference and of a contribution from the spontaneous curvature of the vesicle. The equilibrium shape of the vesicle changes when its environment is modified as these two parameters change. 

In Ref.~\cite{Lee99}, the equilibrium shape of a SOPC-GUV was studied as a function of the external pH, at a fixed internal pH. It was found that increasing the external pH induced shape transitions toward outward curved shapes, whereas lowering it led to inward curved shapes. In light of the ADE model, and given that no visible change of the area or of the volume of the GUV occurred, which implies that $v$ was constant, the authors interpreted this as a variation of $\overline{\Delta a_0}$ with the pH. More precisely, the cause of the shape variations was attributed to a change of the spontaneous curvature of the membrane, under the assumption that the preferred area per lipid was not modified~\cite{Lee99}. Note however that the ADE model does not allow for distinguishing a change of the spontaneous curvature from a change of the preferred area difference.

There is a qualitative link between the membrane response to the static and global pH changes of Ref.~\cite{Lee99} and the dynamical and local pH changes of Refs.~\cite{Khalifat08, Fournier09}: in both cases, the membrane tends to deform outwards when the external pH is increased and inwards when it is decreased. However, the theoretical description of the local and dynamical deformations described in Refs.~\cite{Khalifat08, Fournier09} is far more complex than the interpretation of the results of Ref.~\cite{Lee99}. Indeed, as membrane properties can feature heterogeneities due to the local perturbation, a global theory such as the ADE model is no longer sufficient, and one must develop a local version of the ADE model. In addition, local modifications are intrinsically dynamical, as a local concentration heterogeneity will decay due to diffusion. Hence, a full dynamical description of the membrane is required. 

\paragraph{Membrane dynamics}

Brochard and Lennon~\cite{Brochard75} investigated the dynamical fli\-cker of red blood cells, and showed that it is caused by thermal fluctuations of the membrane shape. This paper introduced a description of the dynamics of a weakly deformed membrane within the Helfrich model~\cite{Helfrich73}. The key is a normal force balance involving the normal elastic force density in the membrane, obtained from the Helfrich Hamiltonian, and the viscous stress exerted by the cytoplasm on the membrane. (For a red blood cell, the viscosity of the cytoplasm is much higher than that of the external fluid, and thus the latter is negligible, which is not the case for a vesicle.)

The description of Ref.~\cite{Brochard75}, adapted to the symmetric case where the fluid above and below a quasi-flat membrane are identical, yields the relaxation rate
\begin{equation}
\gamma=\frac{\kappa q^3}{4\eta}\,, \label{broch}
\end{equation}
for a plane wave with wave vector of norm $q$ deforming a membrane~\cite{Seifert93}. In this formula, $\kappa$ is the Helfrich bending rigidity of the membrane, while $\eta$ is the viscosity of the fluid that surrounds the membrane. Qualitatively, if the membrane deforms, it relaxes due to the bending rigidity $\kappa$, but the relaxation is slowed down by the viscosity $\eta$ of the surrounding fluid, yielding the relaxation rate $\gamma$.

The description of membrane dynamics in Ref.~\cite{Brochard75} was based on the Helfrich model, in which the membrane is treated as a single surface. However, when the membrane curves, one monolayer locally stretches while the other is locally compres\-sed. The resulting density heterogeneities within each monolayer relax by lateral lipid flow, since each monolayer is a two-dimensional fluid. Hence, a more complete description should take into account the coupling between bending and relative compression in membrane dynamics~\cite{Evans92}, and the friction between the two monolayers of the membrane~\cite{Evans88, Merkel89}. 

\begin{figure}[htb]
  \begin{center}
    \includegraphics[width=0.55\columnwidth]{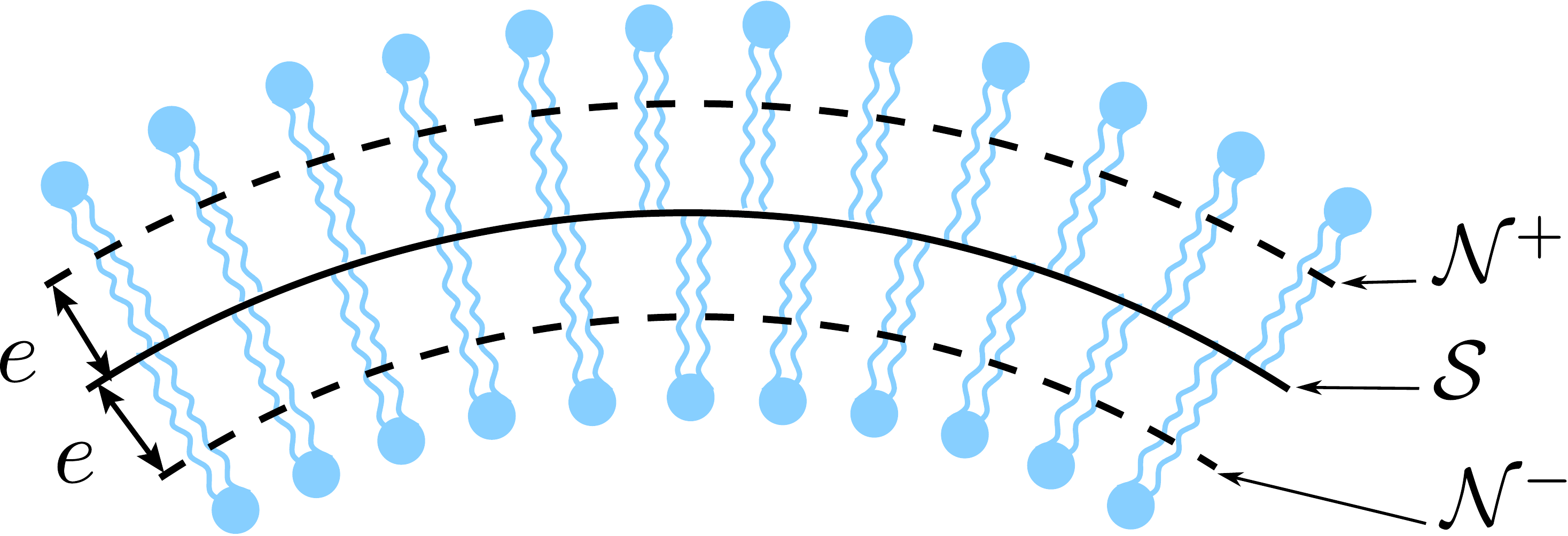}
    \caption{Two-dimensional sketch of a lipid bilayer composed of two identical monolayers. The curvature $c$ and the scaled density $r^{\pm}$ of monolayer $\pm$ are defined on $\mathcal{S}$. The distance between $\mathcal{S}$ and the neutral surface $\mathcal{N}^\pm$ of monolayer $\pm$ is denoted by $e$. If the orientation convention is chosen in such a way that $c<0$ on the drawing, the densities on $\mathcal{N}^\pm$ are $r_n^\pm=r^\pm \pm e c+\mathcal{O}(e^2 c^2)$.}
  \label{figsei}
  \end{center}      
\end{figure}

In order to describe such effects, a membrane model including lipid density heterogeneities is needed. This corresponds to a local version of the ADE model, accounting for the local density in each monolayer. Such a model was presented by Seifert and Langer in Ref.~\cite{Seifert93}. Considering a membrane composed of two identical monolayers, they wrote the following effective Hamiltonian:
\begin{equation}
H = \int_A dA\left[\frac{\kappa}{2}\,c^2+\frac{k}{2}\left(r^++ec\right)^2+\frac{k}{2}\left(r^--ec\right)^2\right],
\label{sei}
\end{equation}
where $A$ is the area of the membrane measured on the midlayer $\mathcal{S}$ between the two monolayers (see Fig.~\ref{figsei}), while $c$ denotes the curvature of the membrane defined on $\mathcal{S}$, and $r^\pm=(\rho^\pm-\rho_0)/\rho_0$ represents the scaled two-dimensional lipid density in monolayer $\pm$, defined on $\mathcal{S}$ too, $\rho_0$ being a reference density. The constant $\kappa$ is the bending rigidity of the membrane, while $k$ is its stretching modulus, and $e$ denotes the distance between $\mathcal{S}$ and the neutral surface $\mathcal{N}^\pm$ of monolayer $\pm$~\cite{Safran,Petrov84} (see Fig.~\ref{figsei}). By definition, the neutral surface of a monolayer is the surface where the stretching and bending modes of the monolayer are decoupled~\cite{Safran}. Since $\mathcal{S}$ and $\mathcal{N}^\pm$ are parallel, the scaled density on $\mathcal{N}^\pm$ reads $r_n^\pm=r\pm e c +\mathcal{O}(e^2 c^2)$, where $\mathcal{O}(e^2 c^2)$ denotes a term of second order in $e c$. Hence, if $H$ is written as a function of $r^\pm_n$ and $c$, it features no coupling between these variables. 

 In this model, the normal force balance still involves the normal elastic force density in the membrane and the normal viscous stress exerted by the fluid above and below the membrane. In addition, a tangential force balance within each monolayer has to be accounted for. It involves the tangential viscous stress of the surrounding fluid, the ``in-plane gradient of the surface pressure'', the viscous stress associated with the two-dimensional lipid flow, and finally, a term of intermonolayer friction. The latter term reads $\mp b(\bm{v}^+-\bm{v}^-)$, where $b$ denotes the intermonolayer friction coefficient, while $\bm{v}^\pm$ is the in-plane velocity of the lipids in monolayer $\pm$. Using these force balances and mass conservation, Seifert and Langer obtained a system of linear differential equations coupling the height of the membrane to the antisymmetric density $r_a=r^+-r^-$~\cite{Seifert93}. Two relaxation rates are obtained from these equations. For modes of sufficiently large wavelengths, they read
\begin{align}
\gamma_1&= \frac{kq^2}{2b}\,,\label{eigenvals-1}\\
\gamma_2&=\frac{\kappa q^3}{4\eta}\,, 
\label{eigenvals-2}
\end{align}
where $q$ denotes the norm of the wave vector considered, while $\eta$ is the viscosity of the fluid that surrounds the membrane. More precisely, the expressions in Eqs.~(\ref{eigenvals-1}) and~(\ref{eigenvals-2}) are valid if $q\ll2\eta k/[b(\kappa+2ke^2)]$, which typically corresponds to wavelengths larger than a few microns. In this regime, $\gamma_2$ corresponds to a pure bending mode, and it is identical as Eq.~(\ref{broch}), while $\gamma_1$ involves intermonolayer friction. Qualitatively, perturbations of the antisymmetric density relax due to the stretching modulus $k$, but this relaxation is slowed down by intermonolayer friction, which is characterized by $b$, yielding $\gamma_1$. Meanwhile, as above, shape deformations relax due to the bending rigidity, and the relaxation is slowed down by the viscosity of the surrounding fluid, yielding $\gamma_2$. 

This model was successfully used to analyze the dynamical fluctuation spectra of bilayer membranes~\cite{Seifert93}. Later, it was extended to describe the spherical budding deformation of a membrane resulting from a local and sudden flip of some lipids from one monolayer to the other in an initially flat membrane~\cite{Sens04}. In this situation, the flip induces a local asymmetry of the density between the two monolayers.

\subsection{Our contribution to the theory for small deformations}

Let us now review our description of the dynamics of a membrane submitted to a local chemical modification in the regime of small deformations. Our description of the bilayer membrane generalizes the local version of the area-difference elasticity membrane model introduced in the seminal work Ref. \cite{Seifert93} and further expanded in Ref.~\cite{Miao02}. We showed that the generic effect of a chemical modification on a monolayer is twofold: both the equilibrium density and the spontaneous curvature are changed. We rigorously expressed the elastic force density in a chemically modified membrane, and derived linear dynamical equations for a chemically modified membrane, which generalize those of Ref.~\cite{Seifert93}. We also performed a calculation of the fraction of the chemically modified lipids in the membrane that arises from the local injection of a reagent above the membrane. In particular, it enabled us to calculate the local pH when an acidic or basic solution is microinjected close to the membrane.

\paragraph{Effective Hamiltonian of a monolayer in the absence of chemical modification}

Our description of the bilayer membrane is based on a local version of the area-difference elasticity membrane model~\cite{Seifert93,Miao02}. In the absence of chemical modification, the local state of monolayer $\pm$ is described by two variables: the total curvature $c$ defined on the membrane midlayer, which is common to both monolayers, and the scaled two-dimensional lipid density $r^\pm=(\rho^\pm-\rho_0)/\rho_0$, defined on the midlayer of the membrane, $\rho_0$ being a reference density. The sign convention for the curvature is chosen in such a way that a spherical vesicle has $c<0$. The outer monolayer corresponds to monolayer $+$, while the inner one corresponds to monolayer $-$.

We write the effective Hamiltonian $f^\pm$ per unit area in monolayer $\pm$ as:
\begin{equation}
f^\pm=\frac{\sigma_0}{2}+\frac{\kappa}{4}c^2\pm\frac{\kappa c_0}{2}c+\frac{k}{2}\left(r^\pm \pm ec\right)^2\,,
\label{fpm0}
\end{equation}
where $\sigma_0$ represents the tension of the bilayer and $\kappa$ its bending modulus, while $k$ is the stretching modulus of a monolayer, and $e$ denotes the distance between the neutral surface~\cite{Safran} of a monolayer and the midsurface of the bilayer. As we assume that the two monolayers of the membrane are identical before the chemical modification, these constants are the same for both monolayers. The two monolayers have opposite spontaneous curvature constants, noted $\mp c_0$, since their lipids are oriented in opposite directions. 

The expression for $f^\pm$ in Eq.~(\ref{fpm0}) corresponds to a general second-order expansion in the small variables $r^\pm$ and $e c$, around the reference state which corresponds to a flat membrane with uniform density $\rho^\pm=\rho_0$. It is valid for small deformations around this reference state: $r^\pm=\mathcal{O}(\epsilon)$ and $ec=\mathcal{O}(\epsilon)$, where $\epsilon$ is a small nondimensional parameter. The construction of the Hamiltonian density Eq.~(\ref{fpm0}) is explained in detail in Ref.~\cite{Bitbol11_stress}. In Ref.~\cite{Bitbol11_stress}, we explicitly showed that the Hamiltonian densities in Eq.~(\ref{fpm0}) give back the ADE model after minimization with respect to $r^\pm$. Hence, our model is a local version of the ADE model, which extends it to the case of inhomogeneous densities.

The effective Hamiltonian $H$ of the bilayer membrane reads
\begin{eqnarray}
H&=&\int_A dA\left(f^++f^-\right)\nonumber\\
&=&\int_A dA\left[\sigma_0+\frac{\kappa}{2}c^2+\frac{k}{2}\left(r^+ + ec\right)^2+\frac{k}{2}\left(r^- - ec\right)^2\right]\,,\,\,\,\,\,\,
\end{eqnarray}
which is consistent with Seifert and Langer's model~\cite{Seifert93} (see Eq.~(\ref{sei})). Note that here, we have taken into account the effect of the membrane tension $\sigma_0$, which was not implemented in Ref.~\cite{Seifert93}.

\paragraph{Chemically modified monolayer}

Let us now focus on the way the membrane effective Hamiltonian is affected by the local chemical modification. We consider that the reagent source, which corresponds to the micropipette tip in an experiment, is localized in the fluid that surrounds the vesicle. Besides, membrane permeation and flip-flop are neglected given their long timescales. Hence, the chemical modification only affects the outer monolayer, i.e., monolayer $+$, and not the inner one. 

Let us denote by $\phi$ the mass fraction of the lipids of the upper monolayer that are chemically modified, and let us assume that the reagent concentration remains small enough to have $\phi=\mathcal{O}(\epsilon)$. We thus have to include this third small variable in our second-order expansion of $f^+$. We obtain (see Ref.~\cite{Bitbol11_stress} for formal details):
\begin{eqnarray}
f^+&=&\frac{\sigma_0}{2}+\sigma_1\phi+\frac{\sigma_2}{2}\phi^2+\tilde\sigma\left(1+r^+\right)\phi\ln\phi
+\frac{\kappa}{4}c^2\nonumber\\&&+\frac{\kappa}{2}\left(c_0+\tilde c_0\phi\right)c+\frac{k}{2}\left(r^++ ec\right)^2\,,
\label{fmod}
\end{eqnarray}
where the constants $\sigma_1$, $\sigma_2$, and $\tilde c_0$ describe the response of the membrane to the chemical modification. These constants depend on the reagent that is injected. 
Besides, the non-analytical mixing entropy term $\tilde\sigma\left(1+r^+\right)\phi\ln\phi$ (see, e.g., Ref.~\cite{Doi_book}) has been added to our second-order expansion. 

\paragraph{Effect of the chemical modification}
Let us now investigate the physical effect of the chemical modification on monolayer~$+$. For this, we study how the equilibrium state of the monolayer described by the Hamiltonian density Eq.~(\ref{fmod}) is affected by the presence of a nonzero $\phi$.

For a homogeneous monolayer (e.g., monolayer $+$) with constant mass, the spontaneous curvature and the equilibrium density can be obtained by minimizing the Hamiltonian per unit mass $f^+/\rho^+$ with respect to $r^+$ and $c$. First, the minimization with respect to $r^+$ gives, to first order in $\epsilon$:
\begin{equation}
r^+_\mathrm{eq}=\frac{\sigma_0}{2\,k}+\frac{\sigma_1\phi}{k}-ec\,.
\label{min1}
\end{equation}
Then, the minimization with respect to $c$ yields to first order, using Eq.~(\ref{min1}):
\begin{equation}
c_\mathrm{eq}=-c_0-\bar c_0 \phi -\frac{\sigma_0 e}{\kappa}\,,
\label{min2}
\end{equation}
where we have introduced 
\begin{equation}
 \bar c_0=\tilde c_0+\frac{2\sigma_1 e}{\kappa}\,. 
\end{equation}
Note that, since we assume that $r^+=\mathcal{O}(\epsilon)$ and $ec=\mathcal{O}(\epsilon)$, we must have $c_0e=\mathcal{O}(\epsilon)$ and $\sigma_0/k=\mathcal{O}(\epsilon)$ for our description to be valid for the values of $r^+$ and $c$ that minimize $f^+/\rho^+$.

The scaled lipid density $r_n^+$ on the neutral surface of the monolayer is related to $r^+$ through $r^+_n=r^++ec$ to first order. This relation arises from the geometry of parallel surfaces~\cite{Safran}, given that the membrane midlayer and the monolayer neutral surface are parallel surfaces separated by a distance $e$. Hence, Eq.~(\ref{min1}) can be rewritten as
\begin{equation}
r_{n,\,\mathrm{eq}}^+=\frac{\sigma_0}{2\,k}+\frac{\sigma_1\phi}{k}\,.
\label{min1b}
\end{equation}
This result is independent of the curvature $c$, contrary to that in Eq.~(\ref{min1}). Indeed, by definition, on the neutral surface, curvature and density are decoupled~\cite{Safran}, while these two variables are coupled on other surfaces. 

Eq.~(\ref{min1b}) shows that, due to the chemical modification, the scaled equilibrium density on the neutral surface of monolayer $+$ is changed by the amount 
\begin{equation}
 \delta r^+_{n,\,\mathrm{eq}}=r^+_{n,\,\mathrm{eq}}(\phi)-r^+_{n,\,\mathrm{eq}}(0)=\frac{\sigma_1\phi}{k} \label{amountr}
\end{equation}
to first order. Besides, Eq.~(\ref{min2}) indicates that the spontaneous curvature of monolayer $+$ is changed by the amount
\begin{equation}
 \delta c_{\mathrm{eq}}=c_{\mathrm{eq}}(\phi)-c_{\mathrm{eq}}(0)=-\bar c_0\phi \label{amountc}
\end{equation}
to first order.

In a nutshell, the effect of the chemical modification (i.e., of $\phi$) on the upper monolayer is twofold. First, the scaled equilibrium density on the neutral surface of the upper monolayer is changed by the amount $\sigma_1\phi/k$ to first order. Second, the spontaneous curvature of the upper monolayer is changed by the amount $-\bar c_0\phi$ to first order. Hence, the constants $\sigma_1$ and $\bar c_0$ describe the linear response of the monolayer equilibrium density and of its spontaneous curvature, respectively, to the chemical modification. This description is generic, but the values of $\sigma_1$ and $\bar c_0$ depend on the chemical modification considered.

\paragraph{Force densities in the membrane}
In Ref.~\cite{Bitbol11_stress}, we derived the complete stress tensor and the resulting elastic force densities in a monolayer  featuring density and composition inhomogeneities to first order in $\epsilon$, using the principle of virtual work. 

Let us focus on small deformations of an infinite flat membrane. Such a description is adapted to practical cases where the distance between the reagent source and the membrane is much smaller than the vesicle radius. It is then convenient to describe the membrane in the Monge gauge by the height $z=h(\bm{r})$, $\bm{r}\in \mathbb{R}^2$, of its midlayer with respect to the reference plane $z=0$. Then, $ec=e\nabla^2 h+\mathcal{O}(\epsilon^2)$ for small deformations such that $\partial_i h=\mathcal{O}(\epsilon)$ and $e\partial_i\partial_j h=\mathcal{O}(\epsilon)$ where $i,j\in\{x,y\}$. Recall that we denote the upper monolayer by~$+$ and the lower one by~$-$.

The force densities in the membrane described by the Hamiltonian densities in Eqs.~(\ref{fpm0}--\ref{fmod}) then read to first order in $\epsilon$~\cite{Bitbol11_stress}:
\begin{align}
\bm{p}_t^+&=-k\,\bm{\nabla}\left(r^++e\nabla^2 h-\frac{\sigma_1}{k}\phi\right)\,,\label{pip_b}\\
\bm{p}_t^-&=-k\,\bm{\nabla}\left(r^--e\nabla^2 h\right)\,,\label{pim_b}\\
p_z&=\sigma_0 \nabla^2 h-\tilde{\kappa}\nabla^4 h-k e\,\nabla^2 r_a-\left(\frac{\kappa \bar{c}_0}{2}-\sigma_1 e\right)\nabla^2\phi\,,
\label{pn_b}
\end{align}
where $\bm{p}_t^\pm$ is the tangential component of the force density in monolayer $\pm$, while $p_z=p_z^++p_z^-$ is the total normal force density in the membrane. In these formulas, we have introduced the antisymmetric scaled density $r_a=r^+-r^-$, and the constant $\tilde\kappa=\kappa+2ke^2$. 

\paragraph{Spontaneous curvature change versus equilibrium density change}
Our general expressions Eqs.~(\ref{pip_b}),~(\ref{pim_b}) and~(\ref{pn_b}) of the force densities in a chemically modified membrane allow for a comparison of the effects produced by an equilibrium density change and by a spontaneous curvature change. Affecting the local spontaneous curvature of a bilayer membrane can yield membrane deformation and budding~\cite{Tsafrir01,Tsafrir03,Staneva05}. Changing asymmetrically the equilibrium density in each monolayer can also result in membrane deformation or budding~\cite{Sens04, Fournier09, Khalifat11}. We considered these two effects at the same time and compared them. 

Eq.~(\ref{pip_b}) shows that the equilibrium density change (i.e., $\sigma_1$) can yield a tangential force density and induce tangential lipid flow. In addition, Eq.~(\ref{pn_b}) shows that both the equilibrium density change and the spontaneous curvature change (i.e., both $\sigma_1$ and $\bar c_0$) can yield a normal force density, and thus a deformation of the membrane. More precisely, changing the spontaneous curvature produces a destabilizing normal force density $\delta p_z^{\,\mathrm{c}}=-\frac{1}{2}\kappa \bar c_0\nabla^2\phi$, while changing the equilibrium density yields $\delta p_z^{\,\mathrm{d}}=\sigma_1 e\nabla^2\phi$. 

\paragraph{Equilibrium state of a chemically modified membrane}
Let us assume for simplicity that a membrane has been chemically modified in such a way that there is a static spatial profile of the mass fraction $\phi$ of the modified lipids in monolayer~$+$. This can occur for instance if the membrane is in chemical equilibrium with a reagent that is continuously injected from a local source, once the stationary profile of reagent concentration in the fluid above the membrane has been reached (see Ref.~\cite{Bitbol13_bba}). 

In the case where only the equilibrium density is affected, which corresponds to $\sigma_1\ne0$ and $\bar c_0=0$, the membrane deformation would vanish once the scaled lipid density on the neutral surface of monolayer $+$ has reached its equilibrium profile $r_{n,\,\mathrm{eq}}^+$, defined in Eq.~(\ref{min1b}). Indeed, the equilibrium condition $p_n=p_i^\pm=0$ of the membrane is then satisfied for $\nabla^2 h=0$, which means that the flat shape is an equilibrium shape. 

In contrast, in the case where only the spontaneous curvature is changed, where $\sigma_1=0$ and $\bar c_0\ne0$, a deformation persists. Indeed, the equilibrium condition $p_n=p_i^\pm=0$ can be satisfied only if $\sigma_0 \nabla^2 h-\kappa\nabla^4 h=\frac{1}{2}\kappa\bar c_0\nabla^2\phi$, which implies that $\nabla^2 h\ne0$ if $\nabla^2\phi\ne0$, so that the plane shape is not an equilibrium shape for a generic inhomogeneous $\phi$. 

Thus, although both mechanisms should lead to membrane deformations, they are not physically equivalent (see  Ref.~\cite{Bitbol13_bba} for a detailed discussion). Generic chemical modifications result both in a change of the spontaneous curvature and in a change of the equilibrium density, so in general both mechanisms are involved.

\paragraph{Relative importance of the two deformation driving forces}
The relative importance of the spontaneous curvature change and of the equilibrium density change can be determined through the ratio 
\begin{equation}
  \left|\frac{\delta p_z^{\,\mathrm{c}}}{\delta p_z^{\,\mathrm{d}}}\right|=\frac{\kappa|\bar c_0|}{2e|\sigma_1|}\label{RATIO}
\end{equation}
of the destabilizing normal force densities associated with each of these two effects.

This ratio depends on the chemical modification at stake. Here, we focus on a chemical modification that affects the lipid headgroups by effectively changing their preferred area. For instance, in the experimental case of an injection of sodium hydroxide on a membrane composed of PC and PS lipids, we expect that the acid-base reaction between the headgroups and the hydroxide ions increases the negative charge of the headgroups and hence the preferred area per lipid headgroup~\cite{Bitbol12_guv}. It is further necessary to resort to a microscopic model to express the modifications of monolayer spontaneous curvature and equilibrium density as a function of the variation of the preferred area per lipid headgroup. 

In Ref.~\cite{Bitbol11_guv}, we presented two very simple such models, a minimal geometrical one and one based on Ref.~\cite{Miao94}. In both cases, we found that the order of magnitude of the ratio in Eq.~(\ref{RATIO}) should be 
\begin{equation}
  \left|\frac{\delta p_z^{\,\mathrm{c}}}{\delta p_z^{\,\mathrm{d}}}\right|\approx \frac{\alpha\kappa}{2e^2 k} \label{RATIO2}
\end{equation}
with $\alpha\approx 1$ a model-dependent constant. Hence, the value of this ratio is about one. In other words, we expect that the effect of the spontaneous curvature change and that of the equilibrium density change have a comparable importance, for a chemical modification of the lipid headgroups which effectively modifies their preferred area. It is thus crucial to take into account both effects in the dynamics of the membrane.

\paragraph{Membrane dynamics}

Using the elastic force densities in Eqs. (\ref{pip_b}--\ref{pn_b}), we describe the dynamics of the membrane to first order in the spirit of Ref.~\cite{Seifert93}. The dynamical equations are best expressed using two-dimensional Fourier transforms of the various fields involved, denoted with hats: for any field $f$ which depends on $\bm{r}$ and on time $t$, $\hat f$ is such that
\begin{equation}
\hat f(\bm{q},t)=\int_{\mathbb{R}^{2}}d\bm{r}\,f(\bm{r},t) e^{-i \bm{q}\cdot\bm{r}}\,.
\end{equation}
The dynamics of the membrane involves the forces specific to the membrane, among which the elastic force densities in the membrane given by Eqs.~(\ref{pip_b}),~(\ref{pim_b}) and~(\ref{pn_b}), but also the viscous stresses exerted by the fluid above and below the membrane. A full description of the hydrodynamics of this fluid is presented in the Appendix of Ref.~\cite{Bitbol13_bba}. Briefly, the velocity field in the fluid above ($+$) and below ($-$) the membrane is caused by the deformation of the membrane and by the lateral flow in the membrane. Mathematically, it is determined by the boundary conditions corresponding to the continuity of velocity at the interface between the fluid and the membrane. Given the short length scales considered, the dynamics of the fluid can be described using Stokes' equation. In addition, the fluid is assumed to be incompressible.

The first dynamical equation that describes the membrane is a balance of forces per unit area acting normally to it~\cite{Seifert93}. It involves the normal elastic force density in the membrane given by Eq.~(\ref{pn_b}) and the normal viscous stresses exerted by the fluid above and below the membrane (see Ref.~\cite{Bitbol13_bba}). It reads:
\begin{equation}
-\left(\sigma_0 \,q^2+\tilde\kappa\, q^4\right)\hat h +k \,e\, q^2\,\hat r_a + \frac{\kappa \,\tilde{c}_0}{2}\,q^2\hat \phi -4\,\eta \,q\,\partial_t \hat h=0\,,
\label{balnorm}
\end{equation}
where $\eta$ denotes the viscosity of the fluid above and below the membrane.

Besides, as each monolayer is a two-dimensional fluid, we write down generalized Stokes equations describing the tangential force balance in each monolayer~\cite{Seifert93}. The first force involved is the density of elastic forces given by Eqs.~(\ref{pip_b}) and (\ref{pim_b}). The second one arises from the viscous stress in the two-dimensional flow of lipids. The third one comes from the viscous stress exerted by the water (see Ref.~\cite{Bitbol13_bba}). The last force that has to be included is the intermonolayer friction~\cite{Evans94}. We thus obtain:
\begin{align}
 -i\,k\,\bm{q}\left(\hat r^+ - e\,q^2\,\hat h - \frac{\sigma_1}{k}\,\hat\phi\right) &-\left(\eta_2\,q^2 +2\,\eta\,q\right)\bm{\hat v}^+ -b\left(\bm{\hat v}^+ - \bm{\hat v}^-\right)=0\,,\label{baltg+}\\
 -i\,k\,\bm{q}\left(\hat r^- + e\,q^2\,\hat h\right) &-\left(\eta_2\,q^2 +2\,\eta \,q\right)\bm{\hat v}^- +b\left(\bm{\hat v}^+ - \bm{\hat v}^-\right)=0\,,\label{baltg-}
\end{align}
where $\bm{v}^\pm$ denotes the velocity in monolayer $\pm$, while $\eta$ is the three-dimensional viscosity of the surrounding fluid, $\eta_2$ is the two-dimensional viscosity of the lipids, and $b$ is the intermonolayer friction coefficient. 

Finally, we use the conservation of mass in each monolayer to first order:
\begin{equation}
\partial_t \hat r^\pm + i\,\bm{q}\cdot\bm{\hat v}^\pm=0\,. \label{massc}
\end{equation}
Considering that each monolayer has a fixed total mass is justified as long as we restrict to timescales much shorter than the flip-flop characteristic time, which is of the order of hours or days in vesicles, depending on the lipid type~\cite{Mouritsen_book}, and which is assumed not to be significantly modified by the local chemical modification. The timescales of the microinjection experiments investigated in our work (typically about 10 seconds, see e.g. Ref.~\cite{Bitbol12_guv}), satisfy this hypothesis.

Combining Eqs.~(\ref{balnorm}),~(\ref{baltg+}),~(\ref{baltg-}) and~(\ref{massc}) yields a system of first-order linear differential equations on $X=(q\,\hat h,\hat r_a)$: 
\begin{equation}
\frac{\partial X}{\partial t}(\bm{q},t)+M(q)\,X(\bm{q},t)=Y(\bm{q},t)\,,
\label{ED}
\end{equation}
where we have introduced the matrix which describes the dynamical response of the membrane~\cite{Seifert93}:
\begin{equation}
M(q)=\left(\begin{array}{cc}
\displaystyle\frac{\sigma_0 q+\tilde\kappa q^3}{4\eta}
&-\displaystyle\frac{keq^2}{4\eta}\\\\
-\displaystyle\frac{keq^3}{b}
&\displaystyle\frac{kq^2}{2b}
\end{array}\right).
\label{ED_defs}
\end{equation}
Here, we have assumed that $\eta_2q^2\ll b$ and $\eta q\ll b$. This is true for all the wave vectors with significant weight in $\hat\phi$, if the modified lipid mass fraction $\phi$ has a smooth profile with a characteristic width larger than 1 $\mu$m.
Indeed, $\eta=10^{-3}\,\mathrm{J\,s/m^3}$ for water, and typically $\eta_2=10^{-9}\,\mathrm{J\,s/m^2}$ and $b=10^9\,\mathrm{J\,s/m^4}$~\cite{Pott02,Shkulipa06}. Besides, the forcing term in Eq.~(\ref{ED}) reads:
\begin{equation}
Y(\bm{q},t)=\left(\begin{array}{c}
\displaystyle\frac{\kappa \tilde c_0 q^2}{8\eta}\hat\phi(\bm{q},t)\\\\
\displaystyle\frac{\sigma_1 q^2}{2 b}\hat\phi(\bm{q},t)
\end{array}\right).
\label{ED_defs_2}
\end{equation}

Eqs.~(\ref{ED}) and (\ref{ED_defs}) show that the membrane deformation is coupled to the antisymmetric density: the symmetry breaking between the monolayers causes the deformation of the membrane. Here, the symmetry breaking is caused by the chemical modification of certain membrane lipids in the external monolayer, i.e., to the presence of $\phi$. And indeed, Eq.~(\ref{ED_defs_2}) shows that the forcing term in Eq.~(\ref{ED}) is proportional to $\hat\phi (\bm{q},t)$.

Note that Eqs.~(\ref{balnorm}),~(\ref{baltg+}),~(\ref{baltg-}) and~(\ref{massc}) also yield a decoupled evolution equation for the symmetric scaled density $r_s=r^-+r^+$:
\begin{equation}
\frac{\partial\hat r_s}{\partial t}=-\frac{kq}{2\eta}
\left(\hat r_s-\displaystyle\frac{\sigma_1}{k}\hat\phi\right)\,,
\label{dynabar}
\end{equation}
where we have assumed that $\eta_2q\ll\eta$, which is true for all the wave vectors with significant weight in $\hat\phi$, if the modified lipid mass fraction $\phi$ has a smooth profile with a characteristic width larger than 1 $\mu$m.

The present theoretical description is general and applies to any local chemical modification of monolayer +, if $\phi$ remains small and if its profile is smooth with a characteristic width larger than 1 $\mu$m. What changes with the nature of the reagent is the value of the constants $\sigma_1$ and $c_0$, which describe the linear response of the membrane to the chemical modification.

\paragraph{Relaxation rates for large wavelengths}

Let us focus on the case where the modified lipid mass fraction $\phi$ has a smooth profile with a characteristic width larger than about 10 $\mu$m. Then, the wave vectors with significant weight in its Fourier transform $\hat\phi$ satisfy $q\lesssim 10^6\,\mathrm{m}^{-1}$. Let us study the eigenvalues of $M(q)$ in this large-wavelength regime.

Membrane tensions are such that $\sigma_0\geq10^{-8}\,\mathrm{N/m}$, and standard values of the other parameters involved in $M(q)$ are $\kappa=10^{-19}\,\mathrm{J}$, $k=0.1\,\mathrm{N/m}$, $e=1$~nm, $b=10^9\,\mathrm{J.s.m}^{-4}$ and $\eta=10^{-3}\,\mathrm{J\,s/m^3}$. Hence, in the large-wavelength regime, we have $q\ll\sqrt{\sigma_0/\tilde\kappa}$, and the eigenvalues of $M(q)$ can be approximated by
\begin{align}
\gamma_1&= \frac{kq^2}{2b}\label{g1}\,,\\
\gamma_2&=\frac{\sigma_0 q}{4\eta}\,. \label{g2}
\end{align}
Indeed, for $q\ll\sqrt{\sigma_0/\tilde\kappa}$, the coefficient $keq^3/b$ in $M(q)$ is much smaller than all the other ones (see Eq.~(\ref{ED_defs})), so that $M(q)$ can be approximated by an upper triangular matrix. 

The eigenvalues of $M(q)$ represent the relaxation rates of a deformation of the membrane. We observe that $\gamma_1$, which involves intermonolayer friction, is identical to Eq.~(\ref{eigenvals-1}) in Seifert and Langer's description~\cite{Seifert93} (see above), while $\gamma_2$ is different, because the effect of tension, which was disregarded in Ref.~\cite{Seifert93}, is actually dominant over that of bending rigidity for large wavelengths.

It is also interesting to compare $\gamma_1$ and $\gamma_2$ to the relaxation rate $\gamma_s$ of the symmetric density, which appears in Eq.~(\ref{dynabar}):
\begin{equation}
\gamma_s=\frac{k q}{2\eta}\,.
\end{equation}
The rupture threshold of a membrane corresponds to a tension of a few mN/m~\cite{Mouritsen_book}. Hence, $\sigma_0\ll k$ for all realistic membrane tensions, which yields $\gamma_s\gg\gamma_2$. Besides, as mentioned above, we have $\eta q\ll b$ for all the wave vectors with significant weight in $\hat\phi$, so that $\gamma_s\gg\gamma_1$. Hence, the relaxation of the symmetric density is much faster than that of the antisymmetric density and of the deformation of the membrane, for all modes in the large-wavelength limit.

\paragraph{Profile of the fraction of the chemically modified lipids}

The time evolution of the deformation of the chemically modified membrane can be determined by solving Eq.~(\ref{ED}). To this end, we first need to determine $\hat \phi (\bm{q},t)$, which is involved in the forcing term $Y (\bm{q},t)$ (see Eqs.~(\ref{ED}--\ref{ED_defs})).

The profile $\phi(\bm{r},t)$ of the mass fraction of the chemically modified lipids in the external monolayer arises from the local reagent concentration increase. We focus on reagents that react reversibly with the membrane lipid headgroups. Besides, we assume that the reaction between the lipids and the reagent is diffusion-controlled (see, e.g., Ref.~\cite{Eigen64}). In other words, the molecular reaction timescales are very small compared to the diffusion timescales. For such a reversible diffusion-controlled chemical reaction, $\phi(\bm{r},t)$ is instantaneously determined by the local reagent concentration on the membrane, which results from the diffusion of the reagent in the fluid above the membrane. 

\begin{figure}[h t b]
\centering
  \includegraphics[width=\columnwidth]{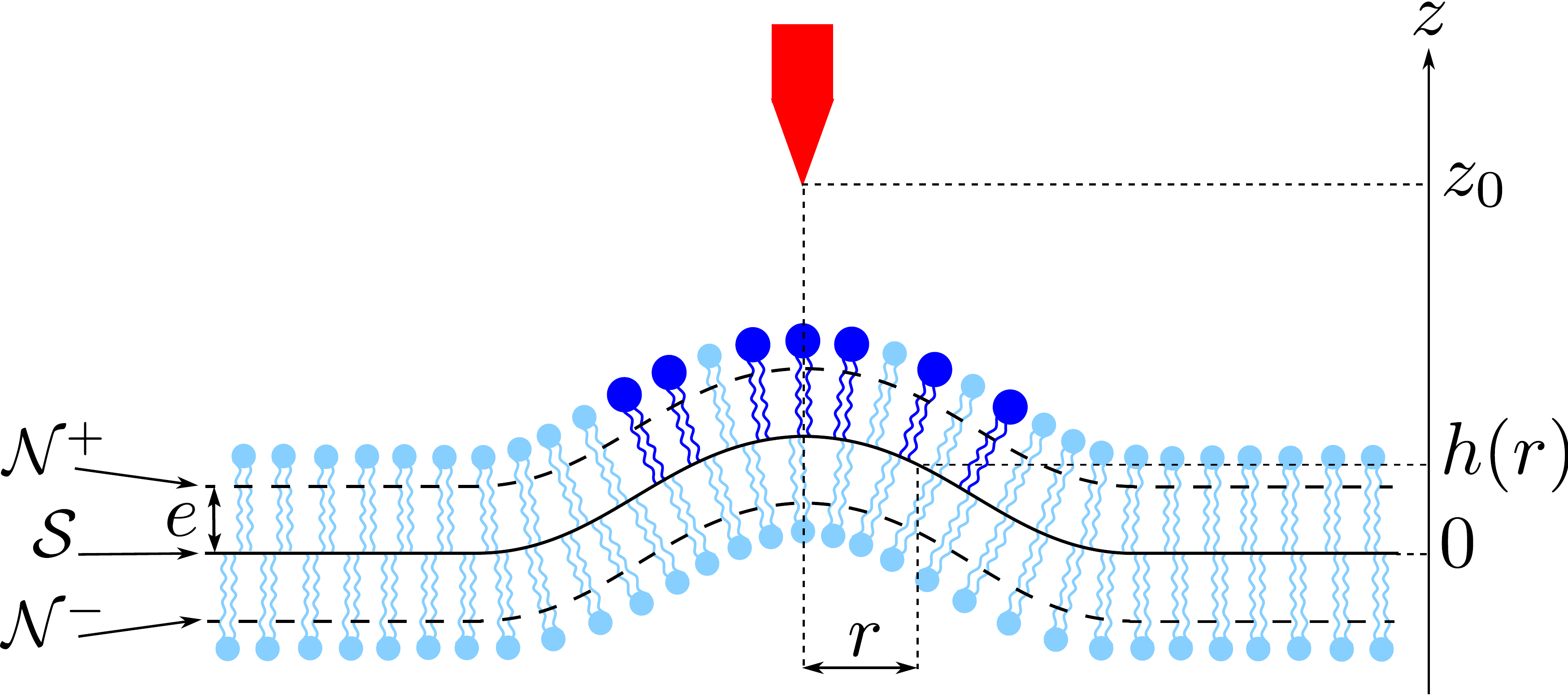}
  \caption{Sketch of the situation described (not to scale). Due to the injection of a reagent from the micropipette (red) standing at $z_0$ above the membrane, some lipids are chemically modified (dark blue) in the external monolayer. The mass fraction of these modified lipids is denoted by $\phi$. The membrane deforms because of this local chemical modification: the shape of its midlayer $\mathcal{S}$ is described by $h(r)$. The variables $c\simeq\nabla^2h$ and $r^\pm$ used in our theory are defined on $\mathcal{S}$, which is at a distance $e$ from the neutral surface $\mathcal{N}^\pm$ of monolayer $\pm$. \textit{Reprinted from Bitbol et al.~\cite{Bitbol12_guv} with permission of the Royal Society of Chemistry.}}
  \label{Situation}
\end{figure}

We consider that the reagent source is localized in $(\bm{r}, z)=(\bm{0}, z_0>0)$, which would represent the position of the micro\-pipette tip in an experiment (see Fig.~\ref{Situation}). The cylindrical symmetry of the problem then implies that the fields involved in our description only depend on $r=|\bm{r}|$. We focus on the regime of small deformations $h(r,t)\ll z_0$, and we work at first order in $h(r,t)/z_0$. Besides, we study the linear regime where $\phi(r,t)$ is proportional to the reagent concentration on the membrane: denoting by $C(r,z,t)$ the reagent concentration field, we have $\phi(r,t)\propto C(r,h(r,t),t)$. To first order in the membrane deformation, this can be simplified into $\phi(r,t)\propto C(r,0,t)$. 

For instance, in the framework of Ref.~\cite{Bitbol12_guv}, where a basic solution was injected close to a vesicle, the acid-base (and complexation) reactions that occur between the lipid headgroups and the injected hydroxide ions are reversible and diffusion-controlled~\cite{Eigen64}. Hence, the local mass fraction $\phi(r,t)$ of the chemically modified lipids is determined by an instantaneous equilibrium with the hydroxide ions that are above the membrane. In addition, in the experimental conditions, the pH on the membrane remains well below the effective pKa of the lipids. Thus, the mass fraction of the chemically modified lipids after the reaction is proportional to the concentration of HO$^-$ ions just above the membrane: $\phi(r,t)\propto C(r,z=0,t)$. 

The field $C$ is determined by the diffusion of the reagent from the local source in the fluid above the membrane. In microinjection experiments, there is also a convective transport of the reagent due to the injection, but the P\'eclet number remains so small that diffusion dominates (see our discussion and experimental results in Ref.~\cite{Bitbol12_guv}). Besides, since the membrane is a surface and $\phi\ll1$, we can neglect the number of reagent molecules that react with the membrane when calculating $C$. Hence, $C$ can be obtained by solving the diffusion equation 
\begin{equation}
\partial_t C-D\nabla^2 C=S\,, 
\end{equation}
where $D$ is the diffusion coefficient of the reagent in the fluid above the membrane, and the source term reads 
\begin{equation}
 S(\bm{r},z,t)=S_0\delta(\bm{r})\delta(z-z_0)F(t).
\end{equation}
This corresponds to an injection flow from the source, with time evolution described by the function $F$. In practice, we will only consider the two following simple cases:\\
1) $F=\theta$, where $\theta$ denotes Heaviside's function: this corresponds to a continuous injection with constant flow, starting at $t=0$ (see Ref.~\cite{Bitbol13_bba}).\\
2) $F=\bm{1}_{[0,T]}$, where $\bm{1}_{[0,T]}$ is the indicator function of the interval $[0,T]$: this corresponds to an injection with constant flow, starting at $t=0$ and stopping at $t=T$ (see Ref.~\cite{Bitbol12_guv}).\\
In addition, the membrane imposes a Neumann boundary condition, which reads $\partial_z C\left(r,h(r,t),t\right)=0$. This relation, which corresponds to a vanishing flux across the membrane, can be simplified to first order into 
\begin{equation}
\partial_z C\left(r,0,t\right)=0\,.
\end{equation}

The solution to this diffusion problem reads
\begin{align}
C\left(r,z,t\right)&=\int_0^{t} \!\!dt' \int_{\mathbb{R}^2} \!\!\!\!d\bm{r'} \int_0^{+\infty}\!\!\!\!\!\!\!\!dz'\,S\left(\bm{r'},z',t'\right)\,G\left(|\bm{r}-\bm{r'}|,z,z',t-t'\right)\nonumber\\
&=S_0\int_0^{t} dt'\,G\left(r,z,z_0,t-t'\right)\,F(t')\,,
\label{cint}
\end{align}
where the causal Green's function $G$ of our diffusion problem can be expressed using the method of images~\cite{Alastuey}:
\begin{equation}
G\left(r,z,z',t\right)=G_\infty \left(r,z-z',t\right)+G_\infty \left(r,z+z',t\right)\,,
\label{G1}
\end{equation}
where we have introduced the infinite-volume causal Green's function of the diffusion equation:
\begin{equation}
G_\infty \left(r,z,t\right)=\frac{\theta(t)}{\left(4\,\pi\,D\,t\right)^{3/2}}\,\exp\left(-\frac{r^2+z^2}{4\,D\,t}\right)\,.
\label{G2}
\end{equation}

For the two simple functions $F$ cited above, combining Eqs.~(\ref{cint}), (\ref{G1}) and (\ref{G2}) provides an analytical expression for $C(r,z,t)$, and for $\hat C (q,z,t)$. Since $\hat\phi(q,t)\propto\hat C\left(q,0,t\right)$, we thus obtain an analytical expression for $\hat\phi$ too.

In the case of a continuous injection (case 1 above), $\hat\phi$ reads
\begin{equation}
\hat\phi_1\left(q,t\right)\propto\mathrm{erf}\left(q\sqrt{Dt}-\frac{z_0}{2\sqrt{Dt}}\right) \frac{\cosh\left(qz_0\right)}{qz_0}- \frac{\sinh\left(qz_0\right)}{qz_0}\,,
\label{phiq2}
\end{equation}
where erf denotes the error function ~\cite{Bitbol13_bba}. As $t\to\infty$, $\hat\phi(q,t)$ converges towards 
\begin{equation}
 \hat\phi_{1,s}(q)\propto \frac{e^{-qz_0}}{qz_0}\,. 
\label{phis}
\end{equation}  

In the case of an injection lasting a time $T$ (case 2 above), we obtain
\begin{equation}
\hat\phi\left(q,t\right)= \hat\phi_1\left(q,t\right)-\theta(t-T)\,\hat\phi_1\left(q,t-T\right)\,,
\label{phiq1}
\end{equation}
where $\hat\phi_1$ is given by Eq.~\ref{phiq2}. 
The value of the proportionality constant in this expression is not crucial for our study since all our dynamical equations are linear: it only affects the deformation and the antisymmetric scaled density by a multiplicative constant.

In Ref.~\cite{Khalifat14}, we used this approach to fully calculate the pH close to a vesicle submitted to a local injection of acid with a constant flow (see Ref.~\cite{Khalifat14}, Supporting Material).

\paragraph{Resolution of the dynamical equations}

Once one has determined $\hat \phi (q,t)$, which is involved in the forcing term $Y(q,t)$ of the dynamical equations (see Eqs.~(\ref{ED}--\ref{ED_defs})), the time evolution of the membrane deformation can be obtained by solving the differential equation Eq.~(\ref{ED}). The square matrix $M(q)$ defined in Eq.~(\ref{ED_defs}) has two real positive and distinct eigenvalues for all $q>0$. Let us call these eigenvalues $\gamma_1$ and $\gamma_2$, and let us introduce the associated eigenvectors $V_1=(v_1,w_1)$ and $V_2=(v_2,w_2)$. For the initial condition $X(q,t=0)=(0,0)$, corresponding to a non-perturbed membrane (i.e., flat and with identical density in the two monolayers), we can write:
\begin{equation}
q\,\hat h(q,t)=\int_0^t ds\,\left[v_1\,e^{-\gamma_1\,t}A(s)+v_2\,e^{-\gamma_2\,t}\,B(s)\right],
\label{qh}
\end{equation}
where $A(t)$ and $B(t)$ are the solutions of the linear system
\begin{equation}
V_1\,A(t)\,e^{-\gamma_1\,t}+V_2\,B(t)\,e^{-\gamma_2\,t}=Y(q,t)\,.
\label{linsys}
\end{equation}

Then, in order to obtain the membrane deformation profile at time $t$, we perform an inverse Fourier transform: 
\begin{equation}
h(r,t)=\frac{1}{2\pi}\int_0^{+\infty} dq\,\,J_0(q\,r)\,\,q\,\hat h(q,t) \,,
\end{equation}
where we have used the cylindrical symmetry of the problem, introducing the Bessel function of the first kind and of zero order $J_0$. Thus, using Eq.~(\ref{qh}), we finally obtain
\begin{equation}
h(r,t)=\frac{1}{2\pi}\int_0^{+\infty} dq\,J_0(q\,r)\int_0^t ds\,\left[v_1\,e^{-\gamma_1\,t}A(s)+v_2\,e^{-\gamma_2\,t}\,B(s)\right].
\label{solution}
\end{equation}
Hence, we can obtain the spatiotemporal evolution of the membrane deformation by carrying out the integrals in Eq.~(\ref{solution}) numerically.

\subsection{Comparison between theory and experiments} 

\paragraph{Experiments in the small deformation regime}
In our experiments of Ref.~\cite{Bitbol12_guv}, the chemical modification of the membrane was achieved by locally delivering a basic solution of NaOH close to the vesicle. This local increase of the pH should affect the headgroups of the phospholipids PS and PC forming the external monolayer of the membrane~\cite{Bitbol11_guv}, as well as the fluorescent marker (when present). 

\begin{figure}[h t b]
  \centering
  \includegraphics[width=\columnwidth]{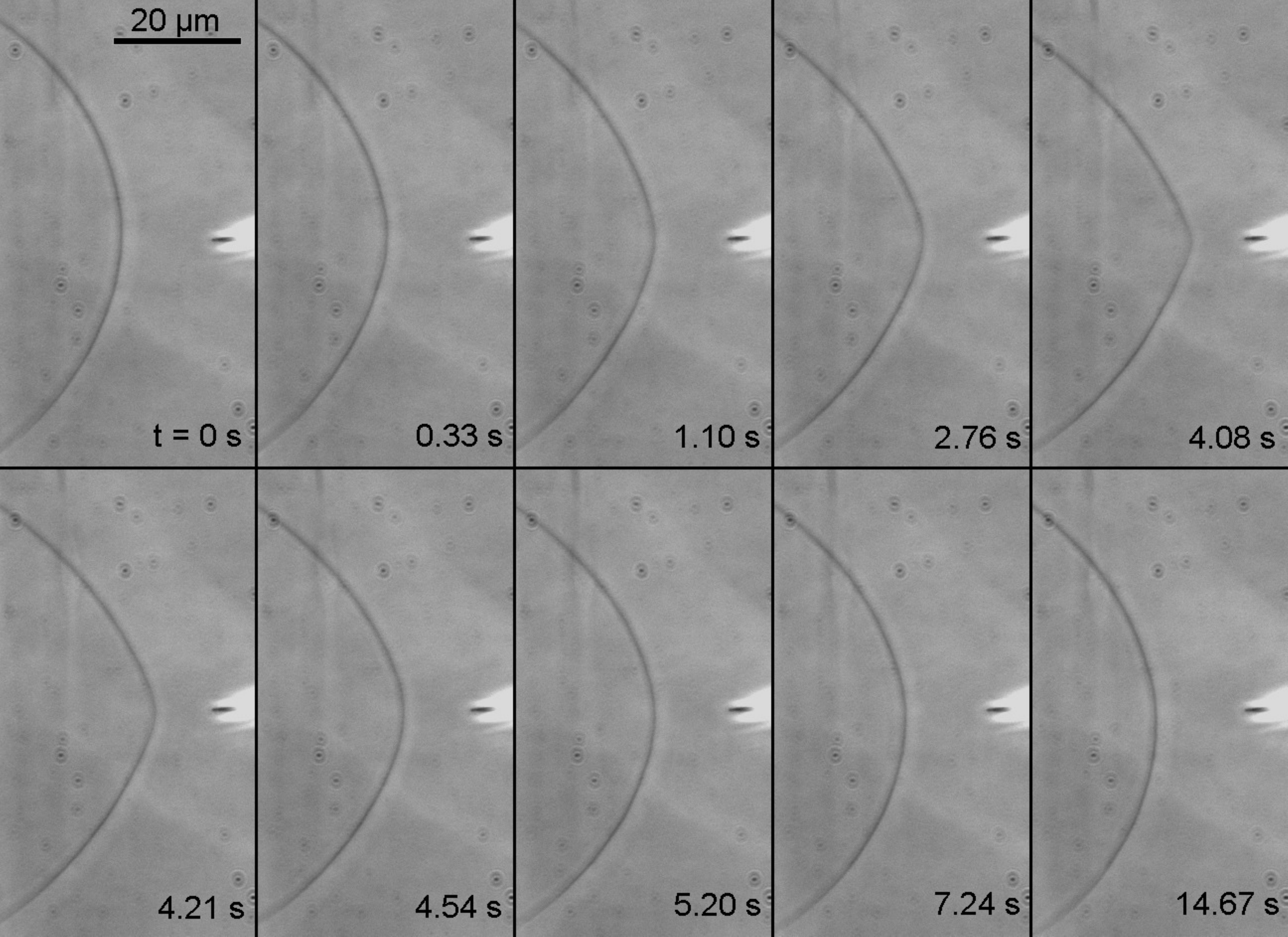}
  \caption{Typical microinjection experiment, lasting $T=4~\mathrm{s}$, observed using phase contrast microscopy. A local modulation of the pH on the vesicle membrane with molar composition EggPC/Brain PS 90:10 (mol/mol) induces a smooth deformation of the vesicle (frames 0.33 to 4.08 s). The deformation is completely reversible when the NaOH delivery is stopped (frames 4.21 s to the end). The single GUV is formed in pure water. \textit{Reprinted from Bitbol et al.~\cite{Bitbol12_guv} with permission of the Royal Society of Chemistry.}
\label{Planche_Phase_Contrast}}
\end{figure}

Fig.~\ref{Planche_Phase_Contrast} shows a typical microinjection experiment. We inject the basic solution during a time $T=4~\mathrm{s}$. 
One can see in Fig.~\ref{Planche_Phase_Contrast} the vesicle before any microinjection (frame 0 s). A smooth local deformation of the vesicle develops toward the pipette during the microinjection (first line of images). Once the injection is stopped, the membrane deformation relaxes (second line of images). This deformation is fully reversible. For the sake of clarity, the deformation presented in Fig.~\ref{Planche_Phase_Contrast} is actually the largest we have studied in Ref.~\cite{Bitbol12_guv}. Indeed, we focused on the regime of small deformations in order to remain in the framework of our linear theory. In particular, it is necessary that the deformation height be much smaller than the distance between the membrane and the micropipette for our theory to be valid. 

\paragraph{Fit to the theory: membrane deformation height}

In order to compare the predictions of our theoretical description to experimental results, we measured the height $H(t)=h(r=0,t)$ of the membrane deformation in front of the micropipette during the microinjection experiments described in the experimental section (see, e.g., Fig.~\ref{Planche_Phase_Contrast}). Several microinjection experiments were carried out on GUVs, with an injection lasting $T=4$~s, and with various distances $z_0$ between the micropipette and the membrane. 

Our experiments of Ref.~\cite{Bitbol12_guv} were performed in the linear regime where $\phi\ll 1$ and $H\ll z_0$, and in the quasi-flat regime $z_0\ll R$, where $R$ is the radius of the GUV. In practice, the radii of our largest GUVs were of order 60 to 80~$\mu$m. We thus focused on values of $z_0$ in the range 10 to 30~$\mu$m. Besides, we adjusted the HO$^-$ concentration for the various $z_0$ in order to obtain small but observable deformations, of order 1 to 5~$\mu$m. Note that in the linear regime, the absolute value of the concentration of the injected solution only affects the deformation by a global proportionality constant~\cite{Bitbol12_guv}. We normalized our experimental data on $H(t)$ by the value of $H(t=4\,\mathrm{s})$, corresponding to the end of the injection, for each experiment. This eliminates the effect of the unknown proportionality constant in our theoretical expression of $\hat \phi$ (see Eqs.~(\ref{phiq1}--\ref{phiq2})), as well as the experimental effect of the different concentrations and of the slightly different micropipette diameters. 

The membrane deformation $H(t)=h(r=0,t)$ predicted theoretically using Eq.~(\ref{solution}) together with Eq.~(\ref{linsys}) and Eqs.~(\ref{phiq1}--\ref{phiq2}) was calculated numerically. This was done in the case of four-second injections, with the values of $z_0$ corresponding to the different experiments, and taking typical values of the membrane constitutive constants: $\kappa=10^{-19}\,\mathrm{J}$, $k=0.1\,\mathrm{N/m}$, $b=10^9\,\mathrm{J.s.m}^{-4}$ and $e=1$~nm, and $\eta=10^{-3}\,\mathrm{J.s.m}^{-3}$ for the viscosity of water. 

In order to solve the dynamical equations, it was also necessary to assign a value to the parameter 
\begin{equation}
\alpha=-\frac{ke\bar c_0}{\sigma_1}\,, \label{defalphaexp}
\end{equation}
which quantifies the importance of the change of the spontaneous curvature relative to the change of the equilibrium density of the external monolayer as a result of the chemical modification (see Eqs.~\ref{RATIO} and~\ref{RATIO2}). As discussed above, rough microscopic lipid models~\cite{Bitbol11_guv} yield $\alpha\approx 1$. In the absence of any experimental measurement of $\alpha$, we took $\alpha=1$ in our calculations. We also checked that the agreement between theory and experiment was not as good for $\alpha=0.1$ and $\alpha=10$ as it is for $\alpha=1$. Interestingly, the value of $\alpha$ cannot be determined from static and global deformations of membranes, since their equilibrium shape in the ADE is determined by a combined quantity involving both spontaneous curvature and equilibrium density modifications (see Section~\ref{context}). However, in Ref.~\cite{Bitbol13_bba}, we demonstrated that $\alpha$ could be deduced from the dynamics of a dynamical membrane deformation in response to a continuous injection, which paves the way for further studies.

\begin{figure}[h t b]
\centering
  \includegraphics[width=0.8\columnwidth]{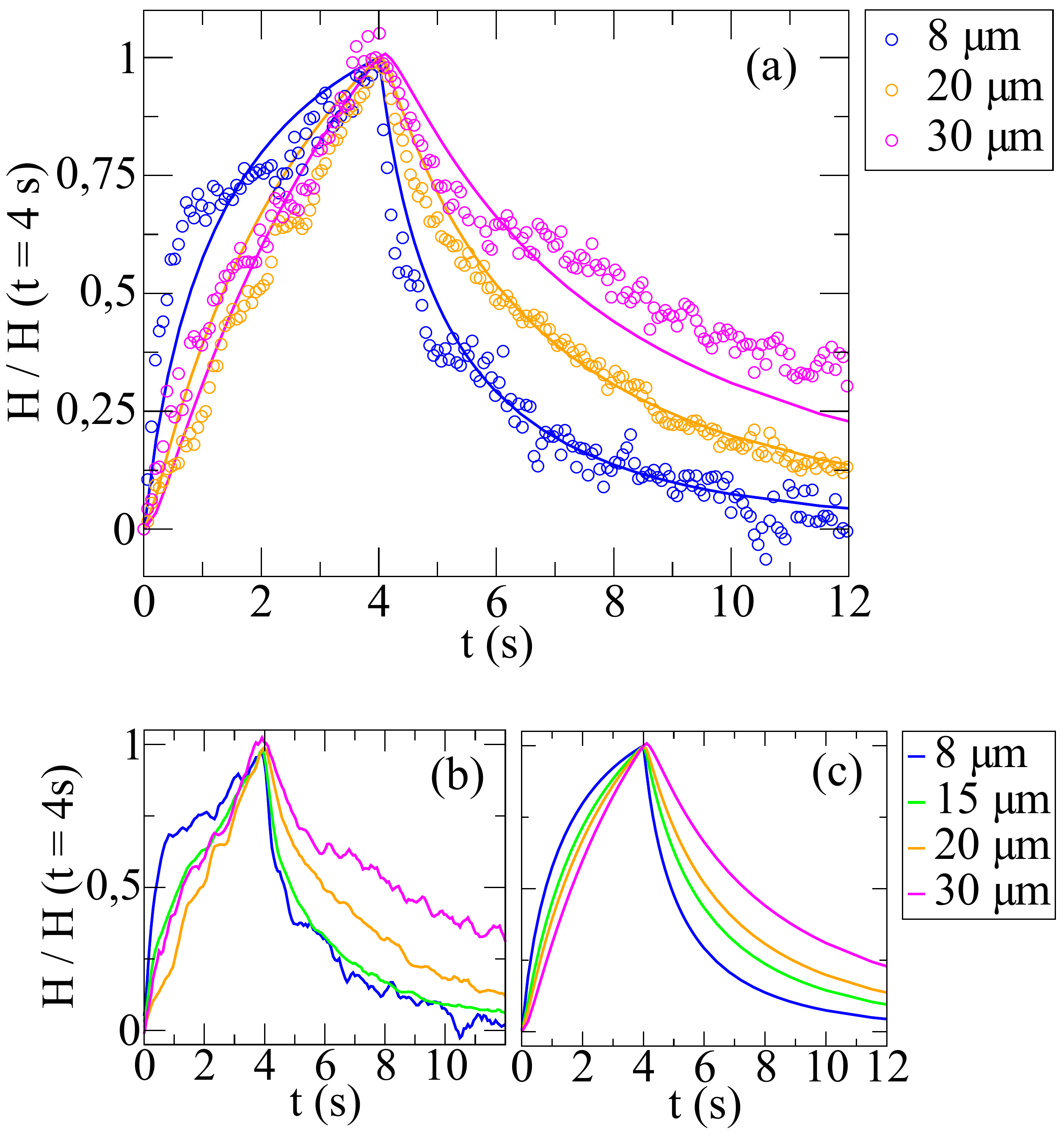}
  \caption{Normalized height $H(t)/H (t=4~\mathrm{s})$ of the membrane deformation, during and after microinjections lasting $T=4$~s. The microinjections were carried out on the same GUV at different distances $z_0$, corresponding to the different colors. (a) Comparison between experimental data and theoretical calculations for three experiments. Dots: raw experimental data (one data point was taken every 57~ms). Solid lines: numerical integration of our dynamical equations with $\sigma_0=1.5 \times 10^{-8}$~N/m. (b) Full set of experimental data; a moving average over 4 successive points was performed to reduce the noise. (c) Full set of numerical data for values of $z_0$ corresponding to those of the experiments. \textit{Reprinted from Bitbol et al.~\cite{Bitbol12_guv} with permission of the Royal Society of Chemistry.}}
  \label{111102}
\end{figure}

\begin{figure}[h t b]
\centering
  \includegraphics[width=0.8\columnwidth]{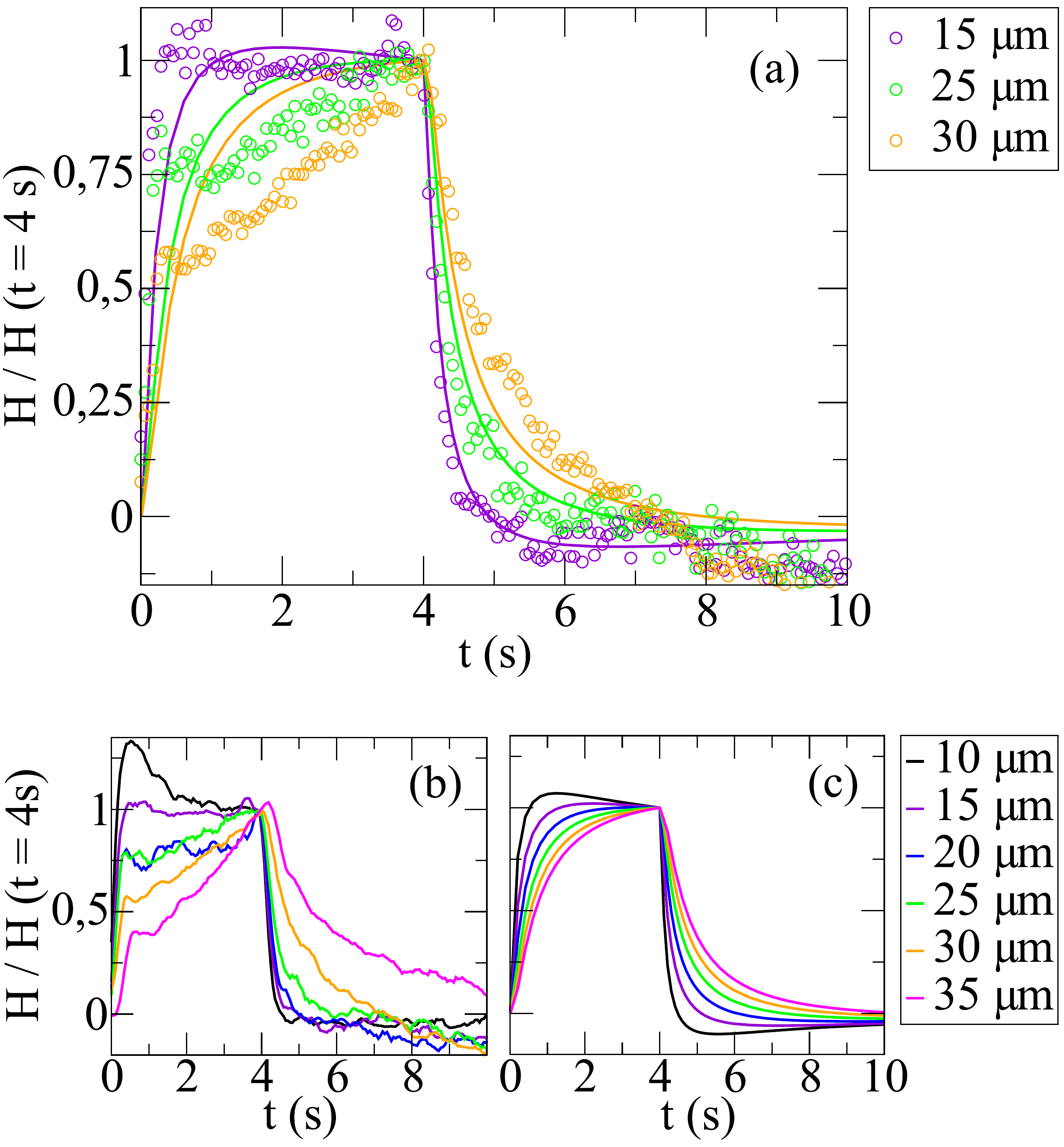}
  \caption{Similar data as on Fig.~\ref{111102}, but for another, less flaccid, vesicle. Here, the numerical integration of our dynamical equations was carried out for $\sigma_0=5 \times 10^{-7}$~N/m. \textit{Reprinted from Bitbol et al.~\cite{Bitbol12_guv} with permission of the Royal Society of Chemistry.}}
  \label{111024}
\end{figure}

Figs.~\ref{111102} and \ref{111024} show the results obtained for several values of $z_0$, on two different vesicles. The only variable membrane parameter was the tension $\sigma_0$. We did not measure this tension during the experiments, but since the vesicles are flaccid, $\sigma_0$ should be in the range $10^{-8}-10^{-6}$~N/m. It was visible that the vesicle corresponding to Fig.~\ref{111102} was more flaccid than that corresponding to Fig.~\ref{111024}. The parameter $\sigma_0$ was adjusted in these two cases, with all the other parameters kept constant at the above-mentioned value. For the vesicle corresponding to Fig.~\ref{111102}, the best match between theory and experiments, i.e., the lowest chi-square, was obtained for $\sigma_0=(1.5\pm0.5) \times 10^{-8}$~N/m, and for the vesicle corresponding to Fig.~\ref{111024}, it was obtained for $\sigma_0=(5\pm1) \times 10^{-7}$~N/m. These values are in the expected range, and the data corresponding to the most flaccid vesicle is best matched by the lowest tension, which is satisfactory.

Figs.~\ref{111102} and \ref{111024} show a good agreement between our experimental data and the results of our theoretical description. In particular, our theory predicts the right timescales of deformation and of relaxation, and also the right variation of these timescales with the distance $z_0$ between the membrane and the micropipette. The increase of the timescales with $z_0$ comes from two different factors. First, the diffusion of the HO$^-$ ions takes longer if $z_0$ is larger. Second, when $z_0$ increases, the width of the modified membrane zone that deforms increases, so that smaller wave vectors have a higher importance in $\hat h (q,t)$. As the relaxation timescales of the membrane, which correspond to the inverse of the eigenvalues of $M$, all increase when $q$ decreases (see Eq.~(\ref{ED_defs})), this yields longer timescales. 

Besides, the timescales involved are shorter if the vesicle is more tense, for instance they are shorter in Fig.~\ref{111024} than in Fig.~\ref{111102}. This can be understood as follows: one of the relaxation timescales of the membrane, which corresponds to the inverse of one of the eigenvalues of $M$ (see Eq.~(\ref{ED_defs})), can be approximated by $4\eta/\sigma_0 q$ for the wave vectors $q$ with largest weight in $\hat h (q)$. This timescale decreases when $\sigma_0$ increases. More qualitatively, a tense membrane will tend to relax faster once it has been deformed.

\subsection{Large deformations: tubulation}

\paragraph{Experiments}
In Ref.~\cite{Khalifat08}, we performed experiments where an acidic solution is microinjected close to a GUV. Local inward deformations of the GUV membrane were observed when it contained negatively charged lipids. In particular, when the membrane composition included cardiolipin, a lipid with four hydrophobic chains that is specific of the inner membrane of mitochondria, dynamical cristae-like invaginations were observed, as shown on Fig.~\ref{Nada} (see section~\ref{minimal}).

\begin{figure}[htb] 
\centering
\includegraphics[width=\columnwidth]{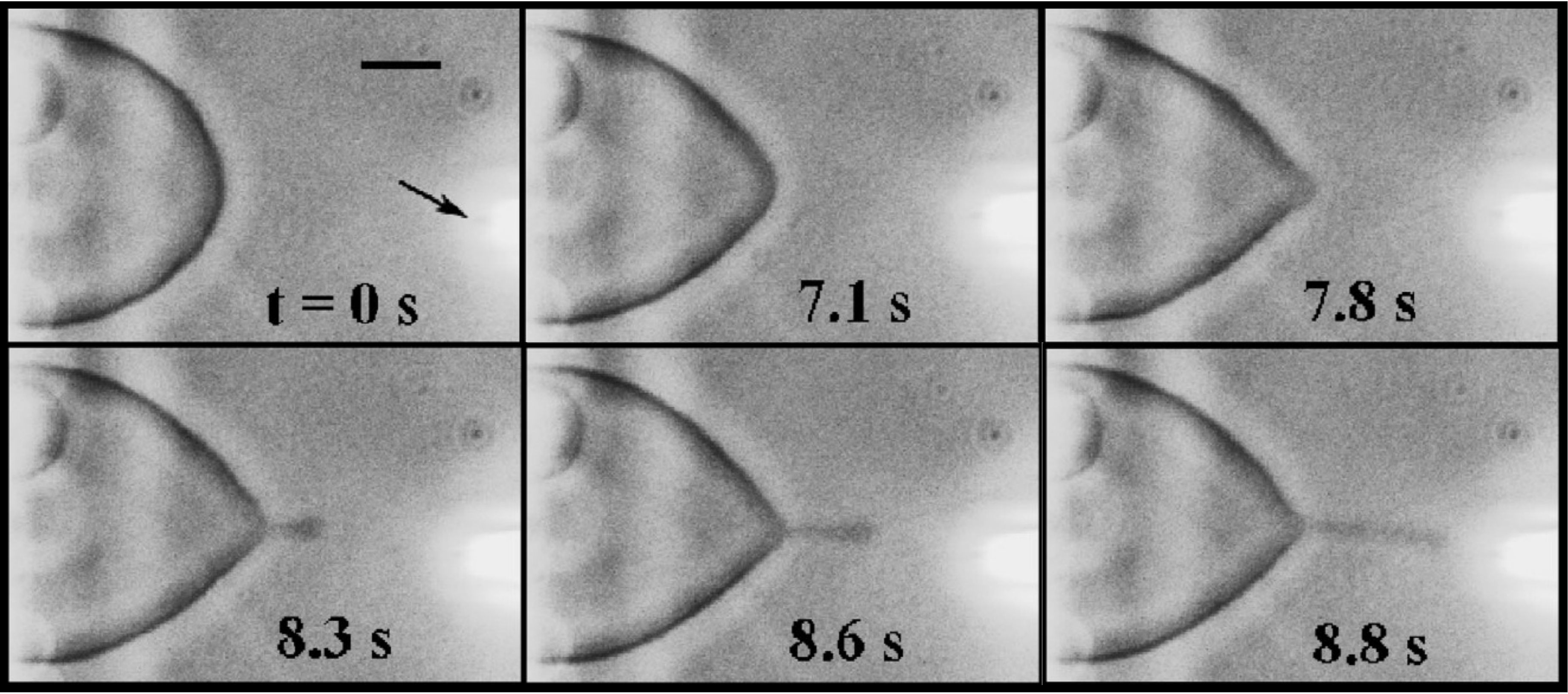}
\caption{Snapshots of a microinjection of a NaOH solution (pH 13) close to a GUV composed of EggPC/Brain PS 90:10 mol/mol in a buffer at pH 7.4. The arrow in the first frame indicates the position of the micropipette tip, and the scale bar represents 10~$\mu$m. The GUV deforms (frames 7.1--7.8~s) and a tubule appears and grows (frames 8.3--8.8~s)~\cite{Fournier09}. \emph{Reprinted from Fournier et al.~\cite{Fournier09} with permission of the American Physical Society.}\label{Figprl}}
\end{figure}

In Ref.~\cite{Kodama2016} (see Fig.~\ref{global}), and as well in Ref.~\cite{Fournier09} (see Fig.~\ref{Figprl}), we described experiments where a basic solution is microinjected close to a GUV. In this case, membrane deformations also occur, but in the opposite direction, i.e., outwards. There can be two phases in the membrane response. First, a smooth outward deformation develops. This small deformation is well described by the above framework. Then, if the concentration of the basic solution is high enough and the distance between the membrane and the micropipette small enough, a very thin tubule appears, and its length grows exponentially in the direction of the micropipette~\cite{Fournier09} (see Fig.~\ref{Figprl}). This second phase cannot be described by our linear theory of small deformations described above. 

\paragraph{Theory and numerical simulations}

In Ref.~\cite{Khalifat08}, we proposed a simple geometric model to describe the inward tubulation (see Fig.~\ref{Nada}) and we obtained, in agreement with the experimental results, long and thin tubes or short and large tubes for respectively strong or weak effect of the acid. We also found that the tenser the vesicle the thinner and longer the tubes. We checked the consistency of our understanding of the invagination process by estimating the orders of magnitude of the energies involved and showed that the inward tubulation is energetically viable.

Ref.~\cite{Fournier09} discussed the threshold of tubule formation, and interpreted the subsequent growth of the tubule as arising from a Marangoni effect. Indeed, the pH gradient on the membrane, which is due to the local injection of the basic solution, can induce a surface tension gradient. We deduced an exponential law for the tubule growth, which matches very well the experimental data~\cite{Fournier09}.

In Ref.~\cite{Khalifat14}, numerical simulations allowed us to further test the consistency of the key ingredients of our theoretical description and of experimental results in the regime of large, nonlinear deformations. Numerical simulations based on a continuum bilayer dynamical model describing the bilayer by its midsurface shape and by a lipid density field for each monolayer were used. The viscoelastic response is determined by the stretching and curvature elasticity, and by the inter-monolayer friction and the membrane interfacial shear viscosity, as in our theoretical description above. However, note that for simplicity, the outer fluid was not explicitly accounted for, neither at the level of its viscous stress, nor at the level of diffusion of the reagent outside the membrane. In addition, any spontaneous curvature effect was neglected, and the focus was on modification of equilibrium density. These hypotheses are discussed in Refs.~\cite{Arroyo09,Rahimi12}. This simplified model was able to capture the nonlinear dynamics observed in our experiments since it allowed for the variation of the lipid densities as a function of time and space as well as for the formation of arbitrary membrane shapes. 

In the simulations, the equilibrium lipid density in the outer monolayer ($+$) was directly increased by a constant amount in a small region of the vesicle for the duration of the injection time. Afterwards, this lipid density modification was gradually decreased with a time-scale commensurate to what we observe in experiments, in order to account for the diffusion of excess H$^+$ after the end of the injection. The simulations featured inward tubulation, in good agreement with our experiments (see section~\ref{PG}).

\section{Biological relevance - mitochondria in health and diseases}
\label{biorel}
The theoretical models reviewed in the previous section allowed us to reveal the physico-chemical mechanisms involving only lipids that initiate membrane deformation and dynamic tubulation,  under local pH gradients. In addition, we elaborated an original artificial lipid-only system presenting dynamic tubular membrane invaginations under acid gradients. We further related our membrane experimental and theoretical studies to the dynamics of mitochondrial inner membrane (IM) morphology and also suggested the underlying role of mitochondrial lipids, especially that of cardiolipin (CL) and its biochemical precursor, phosphatidylglycerol (PG). Our results support the hypothesis of localized bioenergetic transduction as far as it permits to explain how the IM morphology becomes self-maintaining while optimizing the synthesis of biological energy. Recently, a surge in studies on Alzheimer's disease (AD)-related neuronal pathologies focused on Amyloid-$\beta$ (A$\beta$)-induced mitochondrial dysfunctions and morphological alterations. In our work we examined these problems using, for the first time, a bio-mimetic artificial membrane approach.

\subsection{Mitochondria}
\label{mito}
Mitochondria are cellular organelles having a key role in life because biological fuel production, namely ATP synthesis by oxidative phosphorylation, occurs in the mitochondrion~\cite{Voet2006}. Mitochondria have a key role in cell death as well -- they were recently recognized as central regulators of the apoptotic program (the programmed cell death) of eukaryotes~\cite{Bras2005, Heath-Engel2006}.  Several mitochondria-related diseases, such as mitochondrial myopathy and various fatigue and exercise intolerance syndromes, still need cures, thus motivating further study of mitochondrial function. 

\begin{figure}[htb]
	\begin{center}
		\includegraphics[width=0.7\columnwidth]{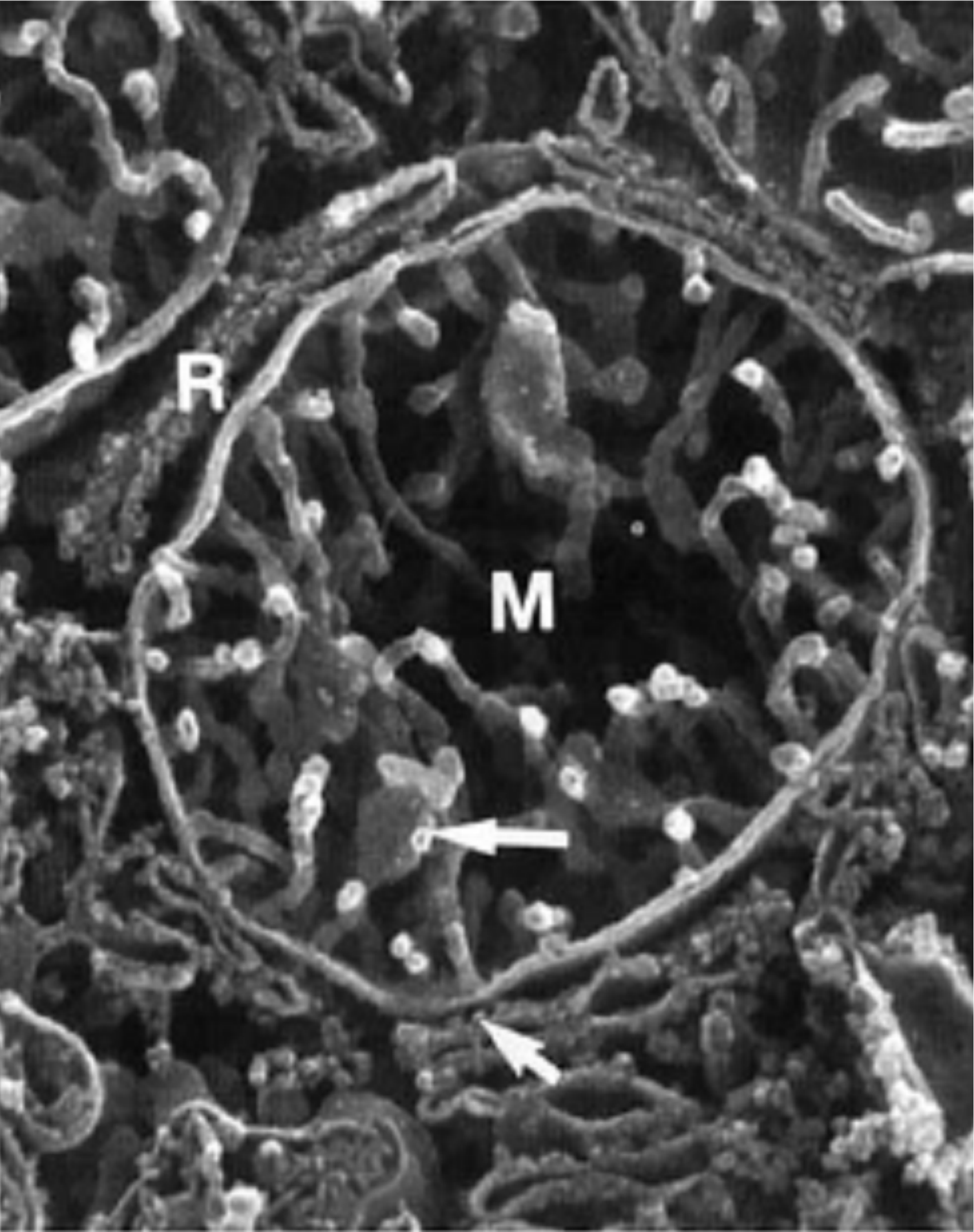}
		\caption{High resolution scanning electron microscopy micrograph of human liver mitochondrion (M), in close proximity to rough endoplasmic reticulum (R). The cristae are tubular, several hundred nanometers long and $\sim$30nm in diameter, x 40 500. \textit{Reprinted from Lea et al.~\cite{Lea1994} with permission of Wiley-Liss.}} 
		\label{mito_fig}
	\end{center}      
\end{figure}

\begin{figure}[htb]
	\begin{center}
		\includegraphics[width=1\columnwidth]{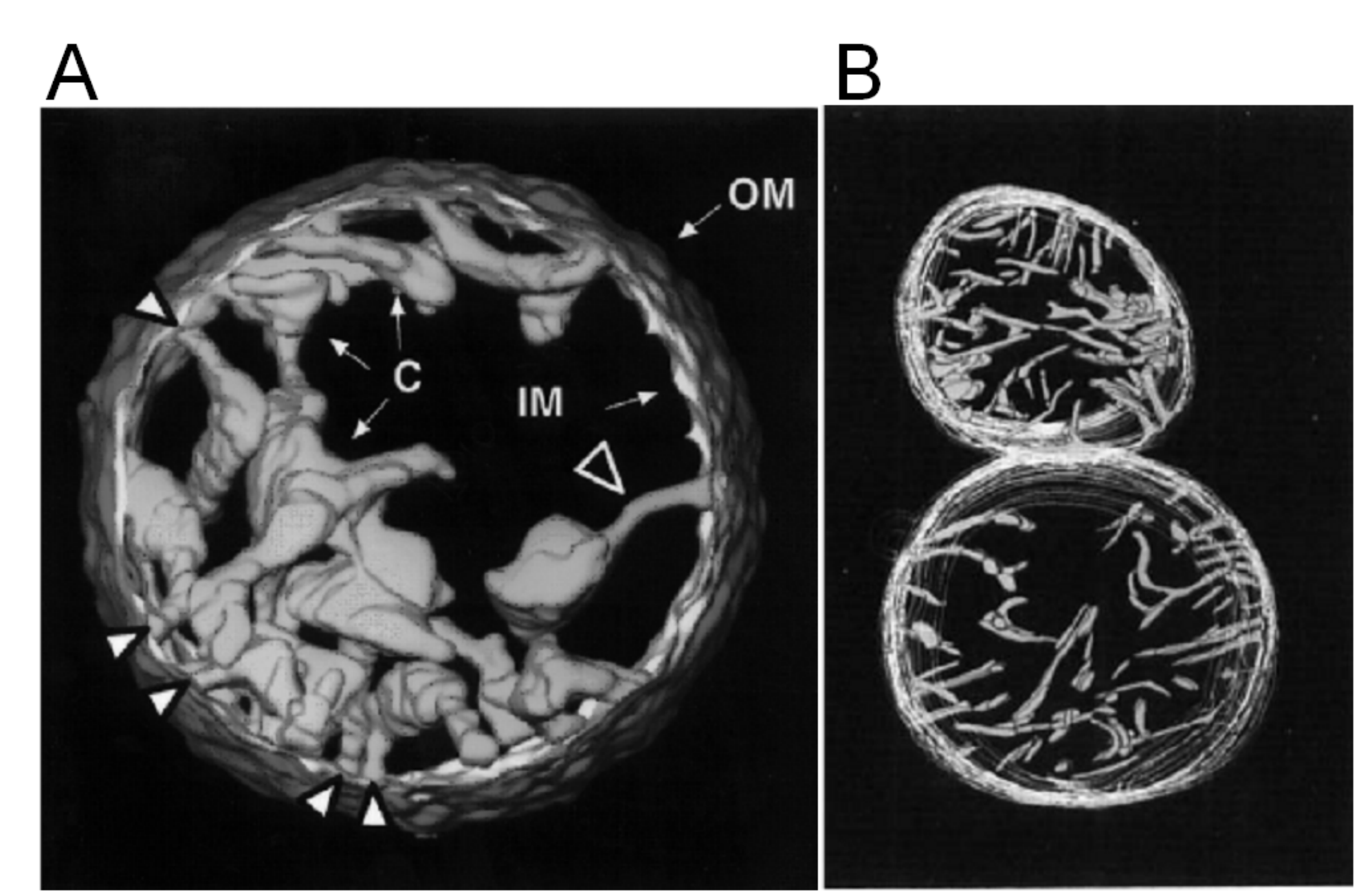}
		\caption{Electron microscopy tomography images of mitochondria: (A) condensed state; (B) orthodox state. OM diameter: (A) 1.5 $\mu$m; (B) lower, 1.2 $\mu$m. (A) \textit{Reprinted from Mannella et al.~\cite{Mannella1998}, with permission of IOS Press.} (B) \textit{Reprinted from Mannella et al.~\cite{Mannella1994} with permission of Wiley-Liss.}}
		\label{OM_IM_fig}
	\end{center}      
\end{figure}

Mitochondria are usually tubular, 1--2 $\mu$m long and 0.1--0.5 $\mu$m large. A mitochondrion is made of two membranes, the outer (OM) and the inner (IM) membranes, separated by the intermembrane space. The IM surrounds a protein-rich medium called the mitochondrial matrix. The IM is made of proteins and lipids 80/20 w/w. The lipid composition of IM is remarkably conserved: it involves large amounts (10--20 mol \% of the total lipid) of a specific lipid, diphosphatidylglycerol, also known as cardiolipin (CL), which is not found in other cellular membranes; phosphatidylcholine (PC) ~40 mol \%; phosphatidylethanolamine (PE) ~30 mol \%; phosphatidylinositol (PI) ~5 mol \%; phosphatidylserine (PS) ~3 mol \%; and sterol ~0.5 mol \%~\cite{Daum1997}. It was shown that CL plays a critical role in mitochondrial function and in diseases including ischemia, hypothyroidism, aging, and heart failure~\cite{Chicco2006}. The OM contains particular proteins (porins) that make it freely permeable to ions and other molecules smaller than 10 kDa, whereas the IM is impermeable even to protons~\cite{Brdiczka1994}. The area of the IM is much larger than the one of the OM. This implies the presence of an important number of IM inward invaginations termed cristae~\cite{Palade1952, Frey2000}. The real 3D structure of mitochondria became accessible only in the 1990s thanks  to the techniques of high resolution scanning electron microscopy~\cite{Lea1994} and of electron microscopy tomography~\cite{Mannella1994, Mannella1998} (for a review, see Mannella~\cite{Mannella2006} and the articles therein). They allow for $\sim$3-nm resolution when looking at an object the size of a mitochondrion and clearly show that the cristae have a predominantly tubular nature (Figs.~\ref{mito_fig} and ~\ref{OM_IM_fig}). For example, the tubular parts of cristae in human liver mitochondria are 30--40 nm in diameter and up to several hundred nanometers long~\cite{Lea1994}. Respiratory chain proteins (i.e. proteins involved in oxidative phosphorylation) reside within the mitochondrial IM, mostly within the cristae compartments~\cite{Dudkina2005, Dudkina2005a, Minauro-Sanmiguel2005, Paumard2002, Schagger2000}, and comprise the electron transport chain proteins, and the ATP synthase complex. During cellular respiration, electron transport chain proteins pump protons from the matrix into the intermembrane space, generating a proton concentration difference ($\Delta$pH) across the IM. The proton concentration difference allows the ATP synthase complex to synthesize ATP~\cite{Logan2006}. The mean bulk pH, measured by pH dependent fluorescence~\cite{Llopis1998}, proved to be $\sim$8 in the matrix, whereas in the intermembrane space it was estimated to be $\sim$7.4 (as in the cytosol) because of rapid diffusion of protons through the numerous pores in the OM. On the other hand, it was suggested that the local pH in the intracristal compartments might be much lower than 7.4 (the bulk pH of the intermembrane space). This hypothesis, called ``Local pH gradient hypothesis", is based on at least three reasons for the existence of local proton concentration gradients within the intracristal compartments: (i), the restricted diffusion of protons between the intracristal and peripheral spaces along the cristae nanotubular structures~\cite{Williams1961, Williams1993, Williams2000, Mannella1997}; (ii), the presence of the negatively charged membrane, and, (iii), the CL capacity as a ``proton trap"~\cite{Haines1983, Haines2002}.

We believe it is now obvious that mitochondria present an astonishing variety of IM morphologies (IM pleomorphism). Moreover, the shape of the inner membrane of a mitochondrion is dynamic, changing in correlation with its functional state~\cite{Heath-Engel2006, Mannella2001} and characteristic for normal (Figs.~\ref{mito_fig},~\ref{OM_IM_fig}) as well as for pathological cases. Normally functioning mitochondria go back and forth from condensed (matrix contracted) to orthodox (matrix expanded) state morphology, depending on their respiratory rate (Fig.~\ref{OM_IM_fig}). Mitochondria carrying out maximum respiratory rate (in the presence of excess ADP and respiratory substrate) present a condensed morphology (Fig.~\ref{OM_IM_fig}A). Their cristae are swollen cisterns or sacs connected to the peripheral part to the IM by narrow tubular segments (cristae junctions) (Fig.~\ref{OM_IM_fig}A, large white arrow). The narrow cristae junctions might have a special function -- restricting the diffusion between intracristal compartments and intermembrane space~\cite{Mannella1997}. The reduction in respiration due to ADP depletion makes mitochondria reverse to the orthodox morphology. The cristae in that state are narrow, flattened or almost tubular (Fig.~\ref{OM_IM_fig}B). On the other hand, the pathologies of mitochondria are correlated with particular morphologies. For example, mitochondrial myopathy is coupled with highly abnormal vesicular cristae. The anaerobic pathogen Cryptosporidium parvum, having a single (relict) mitochondrion and a deficient respiratory chain, presents one of the very few cases of highly folded inner membrane that lacks the typical cristae-like morphology. 

Looking at 3D images of mitochondria, one is logically led to ask how mitochondria function determines IM morphology and, conversely, how the IM morphology can influence mitochondrial function~\cite{Mannella2006, Logan2006, Mannella2006a}. What are the factors that determine IM morphology? The common view has focused on the role of proteins, in particular on those inducing membrane curvature by their cone shape, such as the ATP synthase dimers~\cite{Dudkina2005a}. The role of tBID proteins~\cite{Epand2002, Esposti2001}, mitofilin multimeric complexes~\cite{John2005}, and that of the generated reactive oxygen species as functional factors for the IM morphology has been widely recognized as well~\cite{Bras2005, Heath-Engel2006}. It remains unclear, however, whether the inner membrane morphology and dynamics are fully governed by specific mitochondrial proteins, or whether lipid mediated processes are also involved. In fact, quite limited attention was paid so far to the purely physical, membrane mechanics aspects. When this aspect was considered, the discussion was limited to the simple fact that the IM has a much larger area than the OM, and therefore it is under a simple mechanical (space) constraint that causes it to fold. But simple folds do not suffice to account for the typical cristae-like functioning morphology. On the other hand, several studies (e.g., Ponnuswamy et al.~\cite{Ponnuswamy2005}) suggested some thermodynamic considerations of cristae morphology, but these studies remain partial and far from any dynamics perspective.

Our approach consists in looking for a minimal physicochemical system that models the dynamic morphology of mitochondrial IM. It involves GUV as a model membrane system~\cite{Angelova1986, Angelovaa, Angelovab, Menger1998} and micropipette manipulations for local pH modulation \cite{Angelova1999, Puff2006, Staneva2005, Wick1996}.

\subsection{Mimicking healthy mitochondria}
\label{minimal}

\begin{figure}[htb]
	\begin{center}
		\includegraphics[width=0.8\columnwidth]{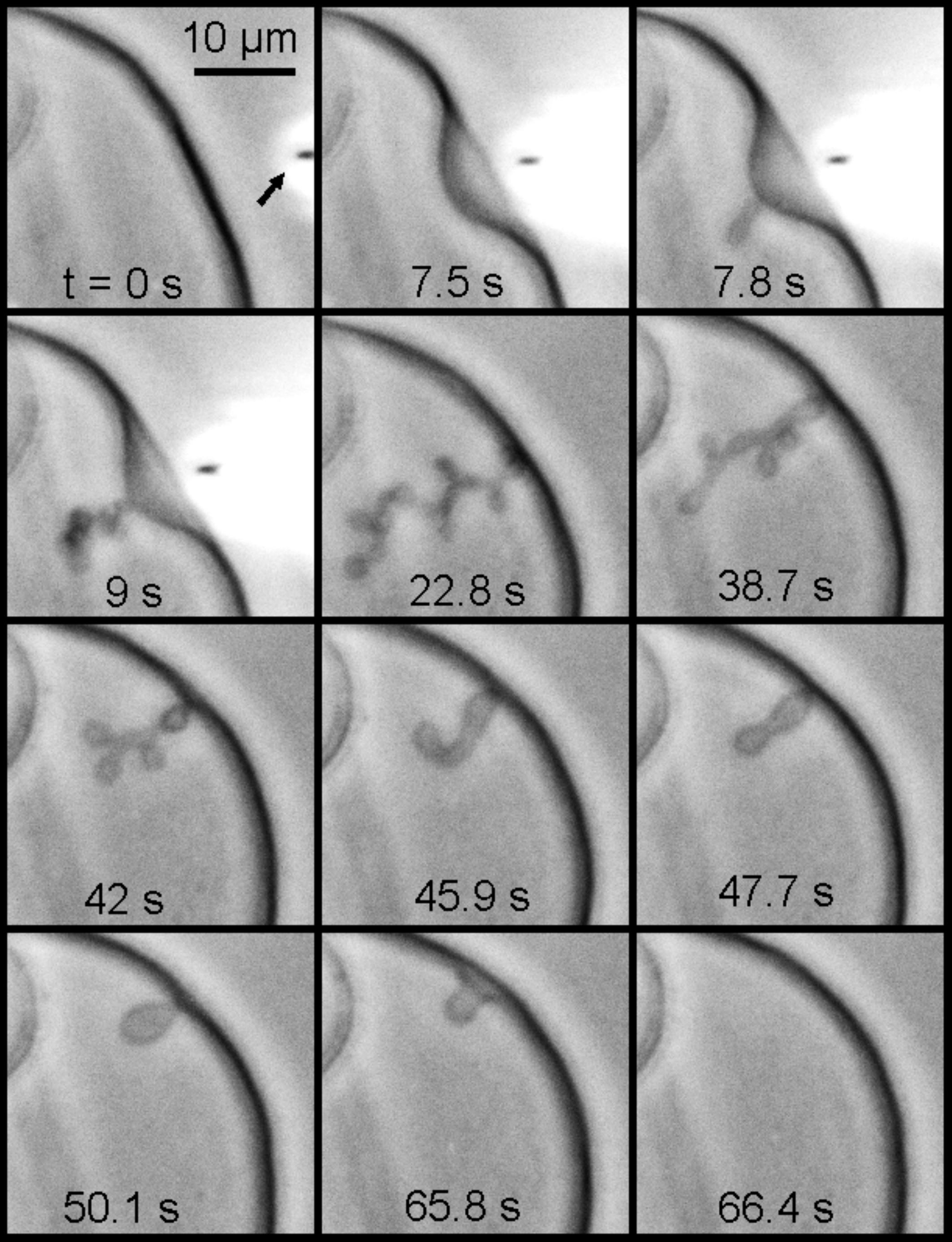}
		\caption{Modulation of local pH gradient at membrane level of a cardiolipin-containing vesicle induces dynamic cristae-like membrane invaginations. The GUV is made of EggPC/EggPE/Heart bovine CL 60:30:10 mol/mol in a buffer at pH 8. The local delivery of HCl (100mM, pH 1.6), which lowers the local pH, is carried out by a micropipette (its position is pointed by the arrow in frame t=0 s). The induced membrane invagination (frame 22.8 s) is completely reversible (frames 38.7--66.4) as far as the acid delivery is stopped. Deflated GUV yield cristae-like morphology with large tubes and vesicular shape segments. \textit{Reprinted from Khalifat et al.~\cite{Khalifat2008} with permission of Elsevier.}}
		\label{Nada}
	\end{center}      
\end{figure}

\begin{figure}[htb]
	\begin{center}
		\includegraphics[width=1\columnwidth]{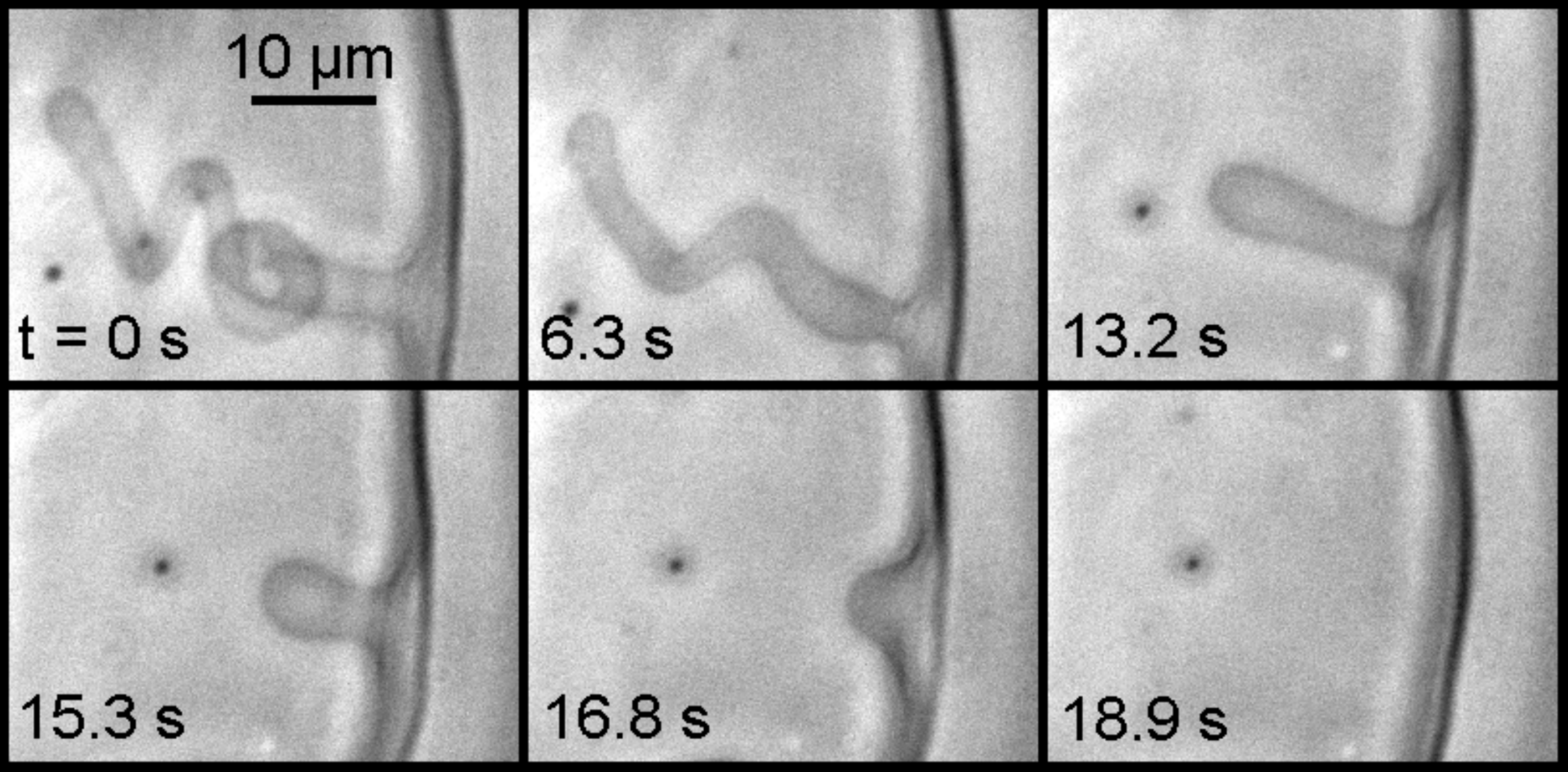}
		\caption{As soon as the acid delivery is turned off, the cristae-like membrane invagination regresses, and it completely disappears when the local pH gradient vanishes. The GUV is made of EggPC/EggPE/Heart bovine CL 60:30:10 mol/mol in a buffer at pH 8. The HCl delivery conditions are as in Fig.~\ref{Nada}. \textit{Reprinted from Khalifat et al.~\cite{Khalifat2008} with permission of Elsevier}.}
		\label{tube}
	\end{center}      
\end{figure}

\paragraph{Design of a minimal model}
The main lipids of mitochondrial IM (presenting 90 mol \% of the total lipid) are CL, PC, and PE. In addition, Kagawa et al.~\cite{Kagawa1973} showed that PC and PE were both required for the reconstitution of vesicles with high 32Pi-ATP exchange activity, and the activity was markedly accelerated by low amounts of CL. Therefore we made our GUV from PC/PE/CL 60:30:10 mol/mol (unless otherwise stated) in a buffer at pH 8 (similar to the bulk pH in the matrix). pH was locally modulated in real time, (lowered and then let to rise to the initial bulk value) by a micropipette delivering (or not) an acid solution outside of the GUV membrane. It is possible to follow directly the induced membrane invagination (Fig.~\ref{Nada}, frame 7.5 s) and the development of a characteristic morphology (Fig.~\ref{Nada}, frames 7.8 s--22.8 s). We identified this morphology as ``cristae-like" as it mimics very well the mitochondrial IM morphology shown by 3D electron microscopy (see Figs.~\ref{mito_fig} and~\ref{OM_IM_fig}). This cristae-like morphology is dynamic, reversible, and modulated by local pH. As soon as the acid delivery is turned off, the cristae-like membrane invagination regresses, and it completely disappears when the local pH gradient vanishes (Fig.~\ref{Nada}, frames 38.7 s--66.4 s, and Fig.~\ref{tube}). We could produce several (e.g., four) cristae-like deformations coexisting dynamically within the same GUV. Invagination shapes, their typical sizes, and characteristic times of relaxation, are dependent on the GUV initial states: either deflated or quasi-spherical. Deflated GUV yield cristae-like morphology with large tubes and vesicular shape segments (Figs.~\ref{Nada} and~\ref{tube}), disappearing within a few tens of seconds after the acid delivery is off. Conversely, the quasi-spherical ones yield morphology with thin and long tubes quite often winding themselves, and being present a long time (up to several tens of minutes) after the acid delivery is turned off. The much longer ``lifetime" of these long thin tubular structures might be a sign of restricted proton backward diffusion along the tube. In consequence, the equilibration of the local pH with that of the bulk might be slowed down due to space factors. Hence, we found experimentally that pH-induced membrane invaginations and cristae-like morphology can be induced just by a sufficient pH gradient.

\begin{figure}[htb]
	\begin{center}
		\includegraphics[width=0.8\columnwidth]{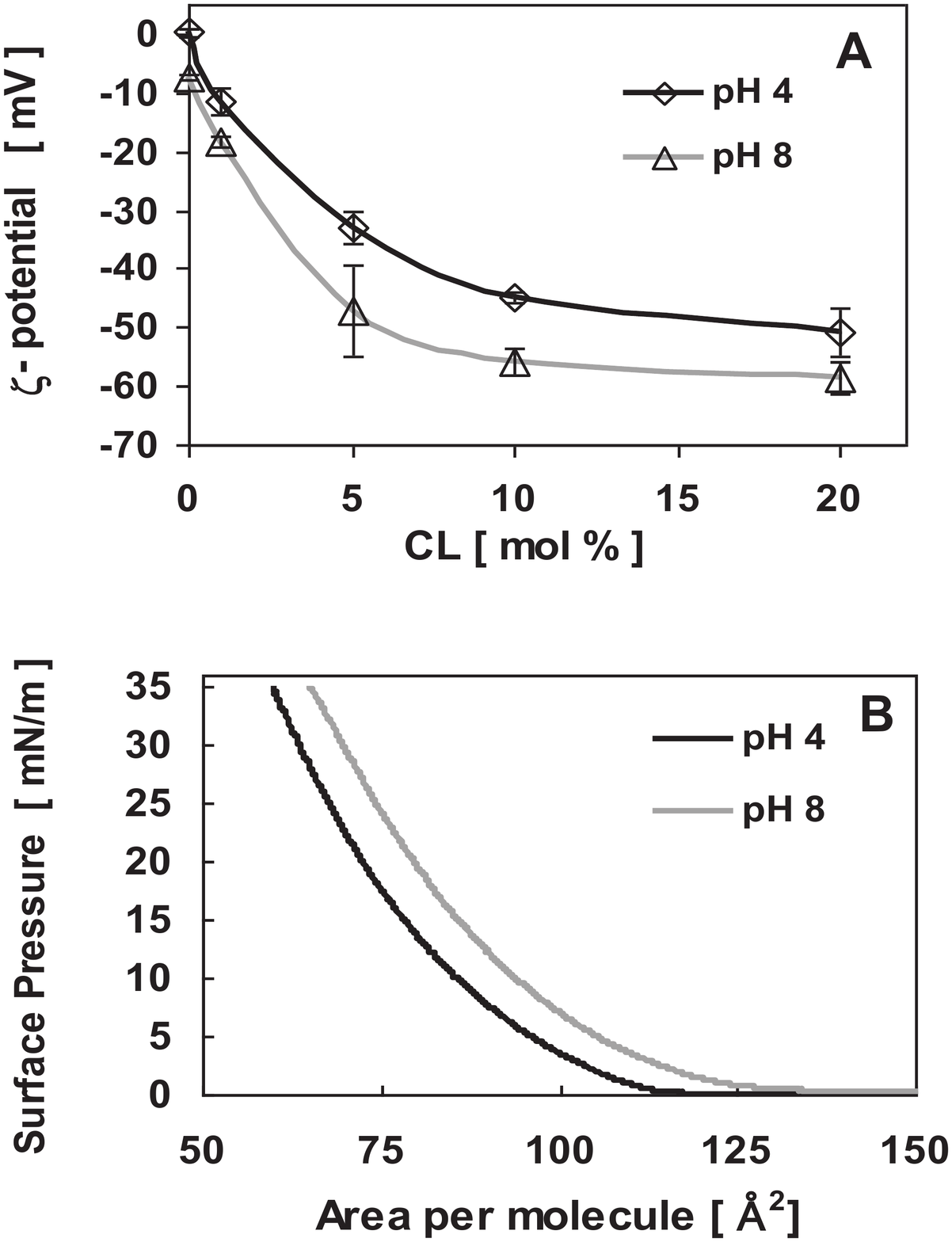}
		\caption{Effect of pH on CL containing vesicles. (A) $\zeta$-potential of LUV made of EggPC/Heart bovine CL 90:10 mol/mol formed in a buffer as a function of CL concentration at pH 8 and pH 4, respectively. The corresponding $\zeta$-potential change for PC/CL is $\sim$20\% (from -56 mV at pH 8 to -45 mV at pH 4). LUV average size was 108 and 122 nm at pH 8 and pH 4, respectively. (B) Mean area per lipid molecule \textit{A} in monolayer made of EggPC/Heart bovine CL 90:10 mol/mol at pH 8 and pH 4, respectively. A decrease of \textit{A} of $\sim$8.6\% (from 70 \AA$^2$ at pH 8 to 64 \AA$^2$ at pH 4) is observed at surface pressure 30 mN/m. The errors in $\pi$ and A were estimated as $\pm$1 mN/m and $\pm$1 \AA$^2$, respectively. \textit{Reprinted from Khalifat et al.~\cite{Khalifat2008} with permission of Elsevier.}
		}
		\label{monolayer}
	\end{center}      
\end{figure}

\paragraph{Invagination mechanism}
Additionally, our experiments, shown in Fig.~\ref{monolayer}, suggest that delivering protons onto CL-containing membrane leads to a partial charge neutralization and area reduction of the exposed monolayer. The induced area mismatch between the inner and outer monolayers of the membrane will create a mechanical stress and induce some deformations.

In this work, we showed  experimentally that proton flow localized at the membrane of CL-containing vesicles induces cristae-like membrane invaginations, and that the morphology of these invaginations is dynamic: one can make the induced invaginations progress or regress by modulating the local pH. This suggests that cristae formation can be accounted for within a purely lipidic minimal model membrane, under the influence of an electrochemical proton gradient. Additionally, we have developed a theoretical description of local membrane deformations attributable to local pH variations (see Sec.~\ref{local} and Refs.~\cite{Bitbol12_guv, Khalifat08,Khalifat11, Fournier09, Bitbol13_bba}). Briefly, the asymmetric pH-induced changes of protonation of the charged headgroups and the resulting changes of the electrostatic interactions between such headgroups promote asymmetric local modifications of the lipid packing. This affects both equilibrium lipid density and monolayer local spontaneous curvature, resulting in a local deformation of the membrane~\cite{Bitbol11_guv, Bitbol13_bba}. Intermonolayer friction plays an important role in the relaxation dynamics of these deformations.

\subsection{Is cardiolipin essential for mitochondrial function?}
\label{PG}
The origin of the shape of mitochondrial inner membrane cristae, as well as that of its coupling to mitochondrial function is the subject of intense research. CL, a particular lipid found almost only in the IM, likely plays an important role because abnormalities in its synthesis usually lead to drastic changes in organization of the IM, including loss of ATP synthesis as well as of the cristae morphology~\cite{Mannella2006}. The crucial function of cardiolipin in mitochondria has been an unresolved question for many (almost a hundred!) years~\cite{Schlame2009}. However, several authors~\cite{Janitor1996, Jiang1997, Chang1998, Jiang2000} reported in the 1990s that the yeast crd1-null mutant, which lacks CL synthase and has no detectable CL but accumulates phosphatidylglycerol (PG), displays normal mitochondrial activity at 30$^{\circ}$C in glucose and nonfermentable carbon sources, although mitochondrial functional defects are apparent under suboptimal substrate conditions and higher temperature. This suggests that PG can specifically substitute for some functions of CL in the IM.

\begin{figure}[htb]
	\begin{center}
		\includegraphics[width=0.95\columnwidth]{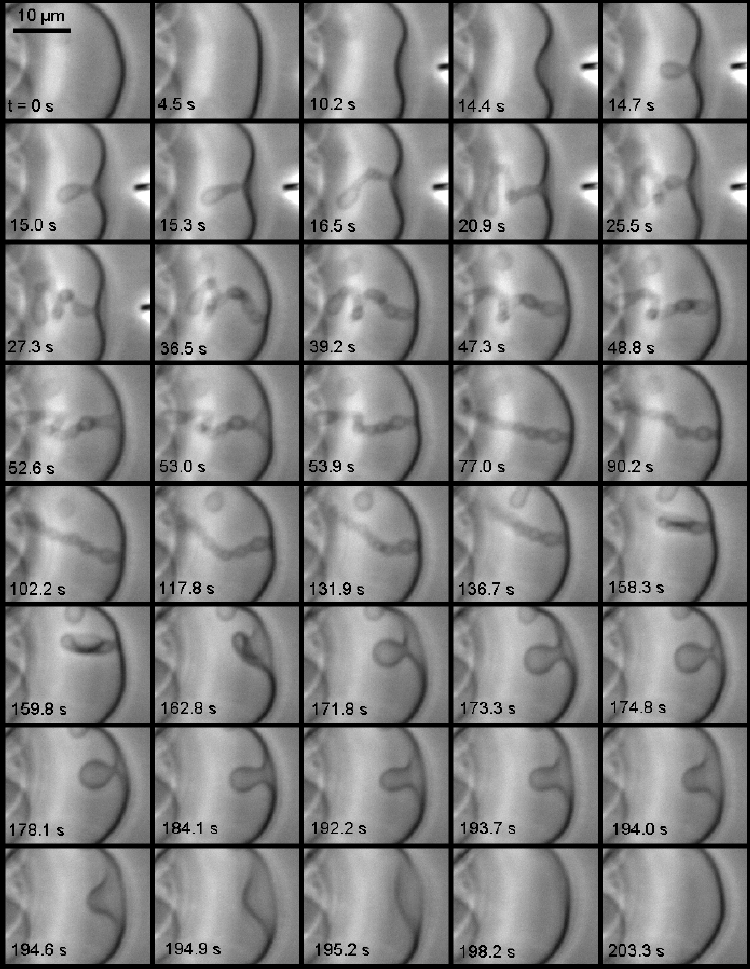}
		\caption{Local membrane invagination triggered on a GUV composed of EggPC/EggPG 90:10 mol/mol by a local pH decrease performed using a micropipette delivering an acidic solution (100 mM HCl, pH 1.6) into the bulk buffer (pH 8, HEPES 0.5 mM, EDTA 0.5mM). The acid microinjection is started at t = 0 and ended at t = 27.3 s. \textit{Reprinted from Khalifat et al.~\cite{Khalifat14} with permission of Elsevier.}}
		\label{PG_fig}
	\end{center}      
\end{figure}

\begin{figure}[htb]
	\begin{center}
		\includegraphics[width=0.8\columnwidth]{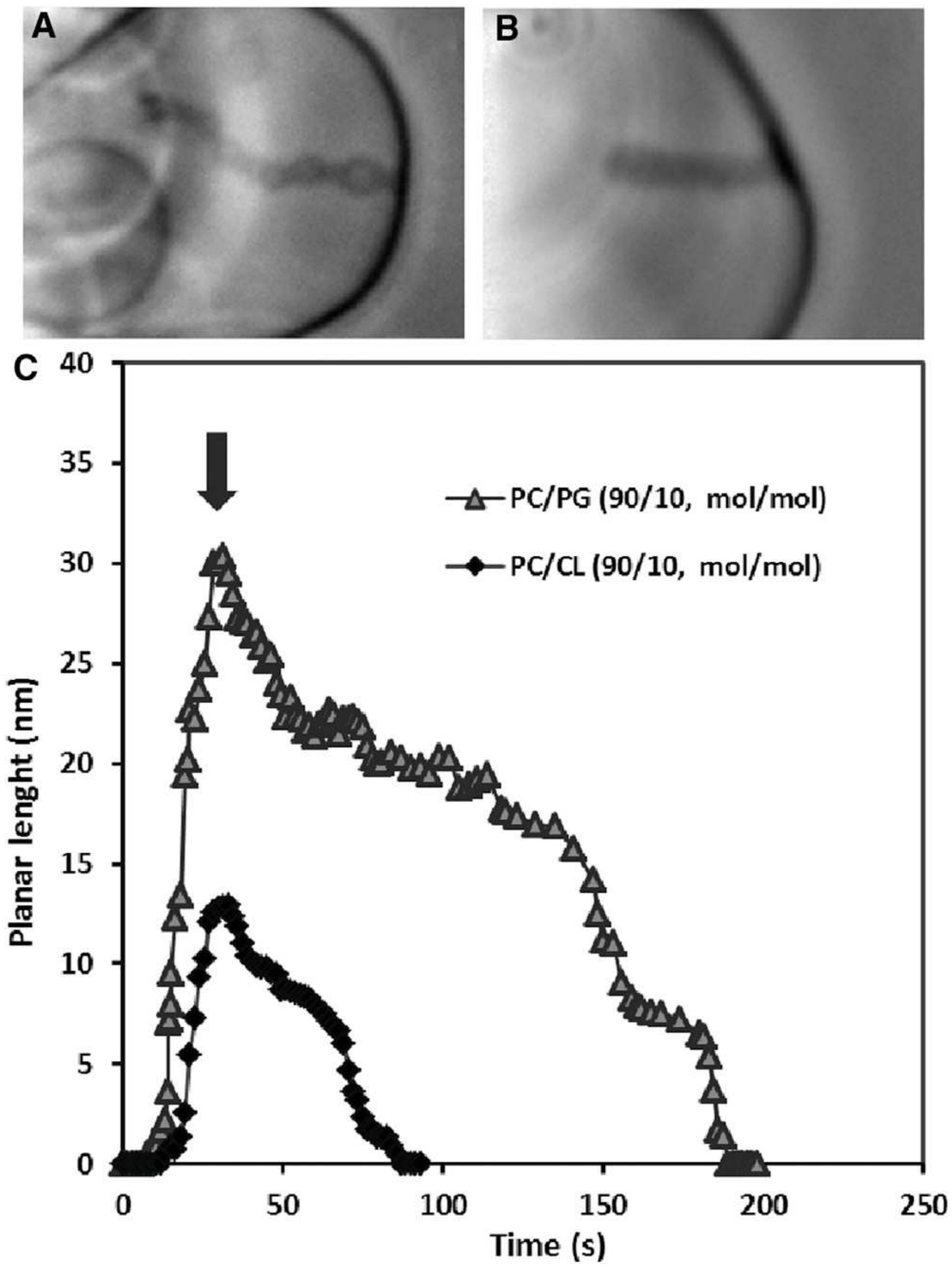}
		\caption{Comparison of the morphologies of tubes formed in GUVs of composition (A) EggPC/EggPG 90:10 mol/mol and (B) EggPC/Heart bovine CL 90:10 mol/mol upon local microinjection with acid. The images correspond to t = 23.7 s and t = 90.2s, respectively, after injection. (C) Comparison of the kinetics of tube growth upon local microinjection with acid for GUVs of composition PC/CL 90:10 mol/mol (diamonds) and PC/PG 90:10 mol/mol (triangles). The apparent planar length of tubes was measured manually on microscopy images using the curvilinear tool of ImageJ. The microinjection starts at t =0 and ends where indicated by a vertical arrow. \textit{Reprinted from Khalifat et al.~\cite{Khalifat14} with permission of Elsevier.}}
		\label{compa}
	\end{center}      
\end{figure}

\begin{figure}[htb]
	\begin{center}
		\includegraphics[width=0.8\columnwidth]{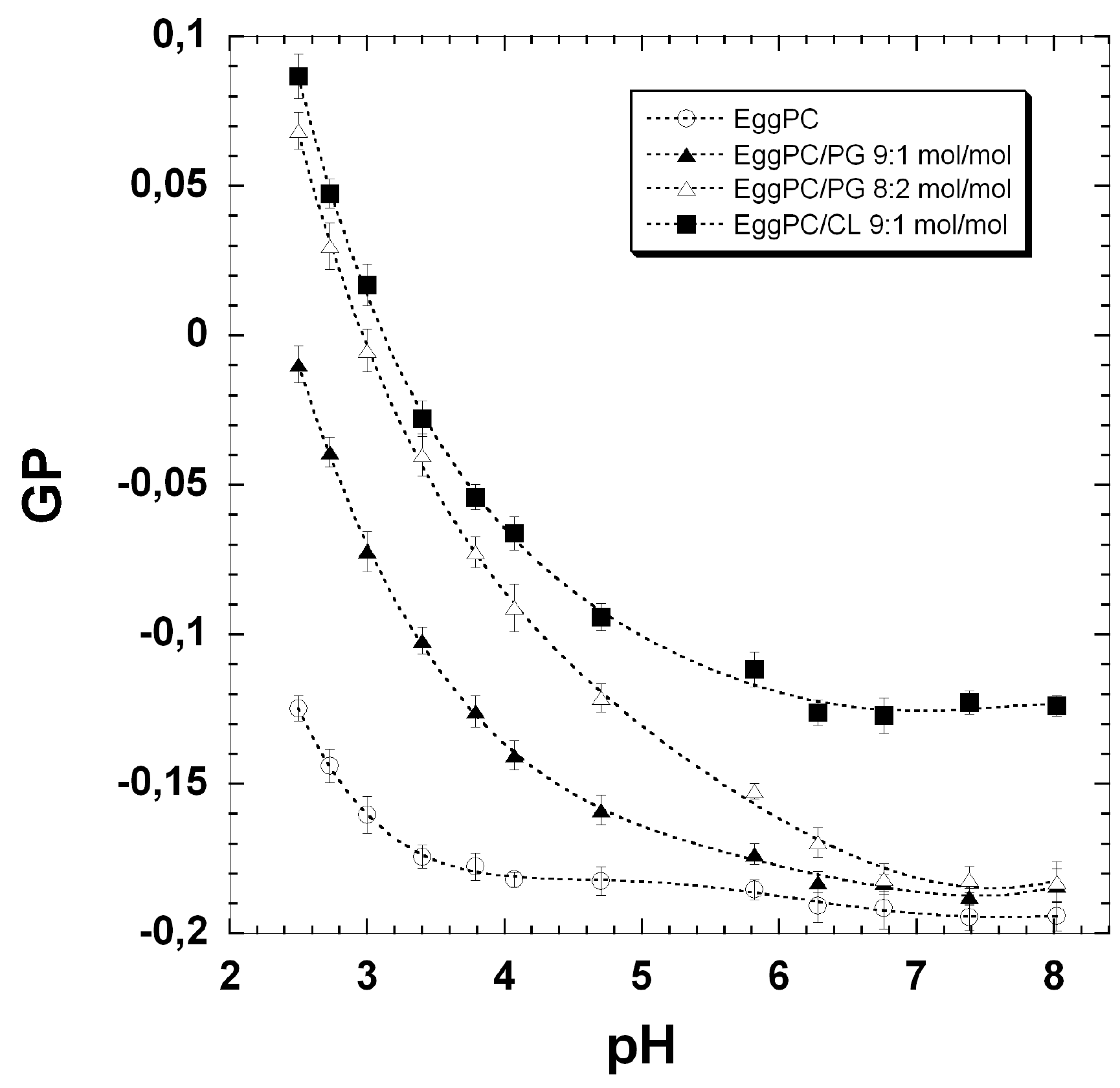}
		\caption{Effect of pH on the general polarization measured from fluorescence emission spectra of Laurdan in LUVs of various compositions in a buffer at 25$^{\circ}$C. The Laurdan/lipid ratio is 1:200 (mol/mol), and the GP is calculated as GP = ($I_{440}$ -- $I_{490}$)/($I_{440}$ + $I_{490}$). \textit{Adapted from Khalifat et al.~\cite{Khalifat14}.}}
		\label{GP}
	\end{center}      
\end{figure}

In our  line of study, we further  refined  the mitochondria model by focusing on yeast mutants that feature CL-deficient but PG-enriched membranes~\cite{Khalifat2014}. Using PC/PG-GUVs, we found that a local pH-induced tubulation also occurs but with a different morphology and dynamics -- faster growth and much slower retraction (see Figs.~\ref{PG_fig} and~\ref{compa}). We  showed that such behavior can be accounted for in the framework of our model, provided that specific properties of PG regarding flip-flop are taken into account. These results provide an explanation of the functionality of the crd1-null yeast mutant. To attain more insight into the mechanism of the local pH induced tubulation in PG-containing bilayers, as compared with CL-containing bilayers, we have studied the effect of pH on lipid packing using the fluorescence of Laurdan incorporated in the membranes of LUVs of similar lipid composition as the GUVs described above. Besides providing useful information about the effect of pH and headgroup protonation state on lipid packing, these data allow one to estimate of the ``apparent pKa" of such protonation phenomena. In bilayers containing charged lipids, and because of the negative surface potential, the H$^+$ concentration in the vicinity of the membrane is likely to be different than in the bulk. Charged lipids, depending on their concentration in the membrane, modulate the pH range for their own protonation. The extent of this modulation also depends on ionic strength~\cite{Ninham1971}. Fig.~\ref{GP} shows the effect of pH on the Laurdan generalized polarization (GP) for various LUVs composition. For PC/CL 90:10, decreasing pH gives rise to an important increase of GP and therefore of lipid packing in the hydrated region of the membrane. This results from the protonation of CL headgroups that decreases the electrostatic repulsion between CL molecules and therefore changes their effective molecular shape as well as their molecular surface area.

We previously showed that such effects are the effective driving force for the pH-induced shape transformations found in CL-containing GUVs~\cite{Khalifat11}. The pH effect on CL bilayers occurs in the bulk pH range of 6 to 2.5, suggesting that this is the actual range for the CL headgroup protonation under such conditions. When PG is substituted for CL in identical molar proportion (PC/PG 90:10), a rather similar effect is observed on GP when pH is lowered, although the overall GP values are higher with CL, because of the well-known ordering effect of the CL backbone~\cite{Lewis2009}. The pH range for the effect is also similar. No such pH-induced packing variation occurs in this range for the zwitterionic phosphatidylcholine, emphasizing its specificity for charged phospholipids. For 10\% PG, the increase in packing occurs between a pH of 7 and 2.5. This is to be compared with a reported intrinsic pK value of 2.9 for PG~\cite{Watts1978}. However, when the proportion of PG in GUVs is doubled to 20\%, the low-pH induced increase in GP is shifted by 0.5 to 1 pH units toward higher pH. This illustrates the mutual influence of PG molecules on their protonation behavior, namely, the higher the PG percentage in the membrane, the more its protonation is anticooperatively hindered because of the unfavorable mutual electrostatic repulsion as well as to the increased membrane charge that attracts H$^+$ toward the surface~\cite{Ninham1971}. Thus the deprotonation of negative lipids in membranes can be strongly shifted or extended to higher pH. These data indicate that, for PG-containing GUVs, a shape-changing mechanism similar to that proposed for CL-containing GUVs is operative~\cite{Khalifat08, Khalifat11}. Indeed, at 10\% PG, a pH decrease from 8 to 4.8 $\pm$ 0.4, identical to that imposed in the GUV experiments, promotes a significant increase in packing. It is therefore likely that such a mechanism contributes to the tubulation observed for both CL- or PG-containing GUVs.

Although the pH-induced tubulation in CL and PG vesicles share one common shape change mechanism, a second mechanism is also operative for PC/PG-GUVs that adds to the surface imbalance, hence the longer tube length and the pearling. This later mechanism is likely also responsible for the slower tube retraction.Interestingly, Cullis and collaborators~\cite{Hope1989, Redelmeier1990} demonstrated that PG can undergo rapid transbilayer diffusion in LUVs through its neutral form even above its pK. In the presence of a pH gradient, the formation of the neutral intermediate is more favorable in the monolayer facing the most acidic solution, so that more PG molecules leave this monolayer toward the other one than the opposite. Furthermore, after monolayer translocation, neutral PG molecules are bound to become quickly deprotonated in the new basic environment, making the reverse motion very unlikely. In short, protonated PG becomes neutral and can easily flip toward the higher pH leaflet driven by mixing entropy differences, but then by losing its proton in the new environment the reverse translocation is kinetically blocked. This results in a net transport of PG toward the monolayer and an asymmetric distribution of PG, with PG accumulating in the monolayer facing the most basic solution. No similar transbilayer diffusion is found for CL~\cite{Hope1989}. This difference can be explained by the smaller size of the PG headgroup compared with the bulky CL headgroup and by that fact that PG can be brought to neutral form by binding of a single H$^+$ whereas CL requires binding of 2 H$^+$, a much rarer event. Considering the well-known relation between phospholipid asymmetric flip-flop and membrane curvature~\cite{Bitbol12_guv,Seigneuret01061984, Farge1992}, we propose that inward PG flip-flop occurs in the area exposed to acid in GUV experiments. Since the external surface of this area is more acidic, lowered to a pH close to 4, this generates a net inward transport of PG from the outer to the inner monolayer thereby creating an asymmetric distribution of PG. Such a process leads to a decrease of the outer monolayer density and an increase in the inner monolayer density and therefore to an additional density mismatch between the two monolayers and therefore to a further bilayer imbalance. This adds up to the direct mechanism involving electrostatic repulsion and modification of equilibrium density mismatch/spontaneous curvature. Interestingly, global shape changes of PG-containing GUVs upon bulk external pH acidification have already been reported~\cite{Seigneuret01061984, Mui1995}.

\begin{figure}[htb]
	\begin{center}
		\includegraphics[width=0.85\columnwidth]{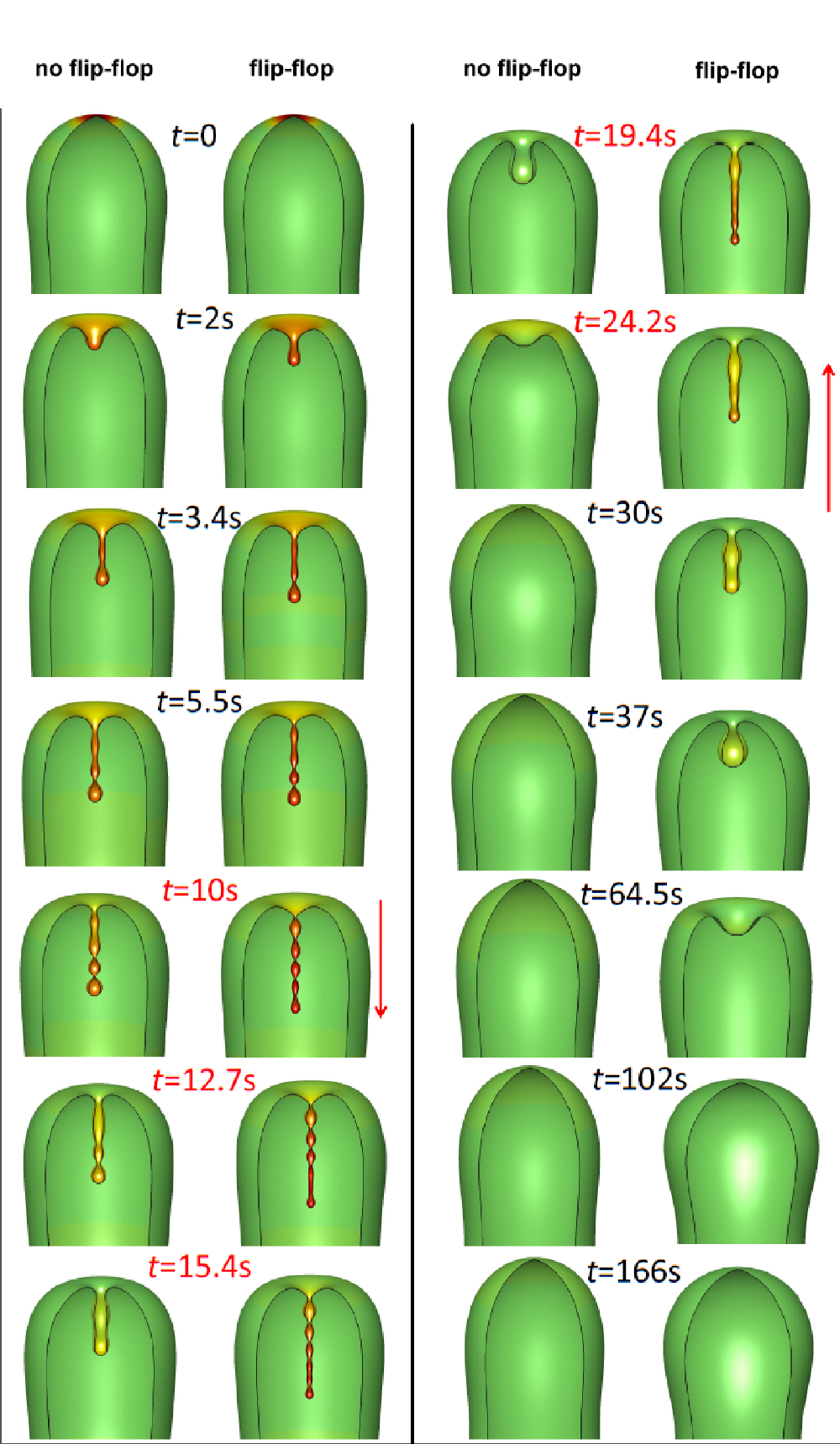}
		\caption{Results of simulation of tube formation after local acid injection on a GUV in the case of direct density change, corresponding to CL-containing GUVs (\textit{first and third columns}), and in the case of direct density change and flip-flop, corresponding to PG-containing GUVs (\textit{second and fourth columns}). The color maps represent the difference between the monolayer lipid densities, where red corresponds to the maximal value and green to 0. The time values and arrows in red correspond to the time period where the acid diffuses and therefore the lipid density modification exponentially decreases. \textit{Reprinted from Khalifat et al.~\cite{Khalifat14} with permission of Elsevier.}}
		\label{simul_fig}
	\end{center}      
\end{figure}

\begin{figure}[htb]
	\begin{center}
		\includegraphics[width=0.8\columnwidth]{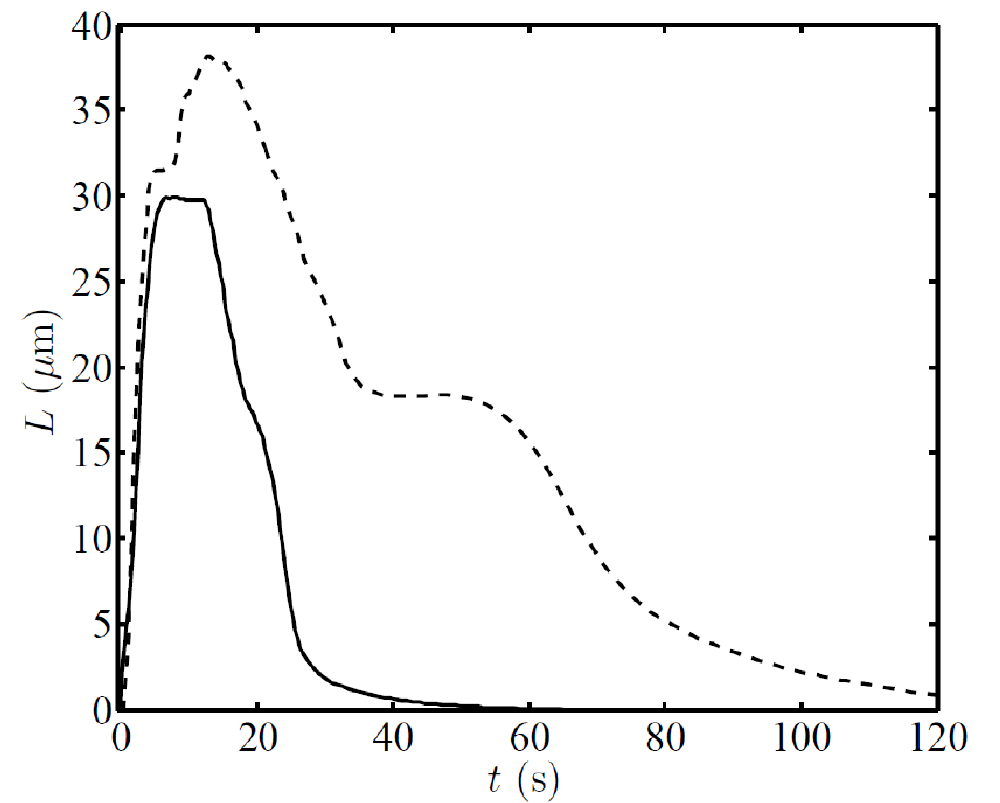}
		\caption{Dynamics of tube length growth derived from simulations and considering either only a direct equilibrium density change of the outer monolayer (\textit{solid line}), corresponding to CL-containing GUVs, or a direct equilibrium density change and transbilayer sink/source flip-flop (\textit{dashed line}), corresponding to PG-containing GUVs. 
			 \textit{Reprinted from Khalifat et al.~\cite{Khalifat14} with permission of Elsevier.}}
		\label{length}
	\end{center}      
\end{figure}

Comparison of the simulations of direct lipid density modification alone (corresponding to CL) and direct lipid density modification plus flip-flop (corresponding to PG) is shown on Fig.~\ref{simul_fig}. The introduction of flip-flop promotes a more rapid and intense tubulation with enhanced pearling. Interestingly, whereas tubulation because of direct density modification reaches a maximum at the time of end of injection, tubulation further increases afterward when flip-flop is taken into account. This can be explained by the fact that additional flip-flop keeps occurring until the pH gradient is dissipated. Fig.~\ref{length} shows the kinetics of growth and retraction of the tube corresponding to the simulation shown in Fig.~\ref{simul_fig}. The tube length and kinetics both in the case of a direct equilibrium density change (corresponding to CL-containing GUVs) and in the case of a direct equilibrium density change plus flip-flop (corresponding to PG-containing GUVs), although not identical to experiments, occur in similar time frames and with similar relative timescales (factor of four in relaxation times).

The present study documents two possible mechanisms involved in pH gradient-induced shape changes of charged lipid bilayers: (i), direct phospholipid packing (equilibrium density and spontaneous curvature) modifications; and (ii), asymmetric flip-flop (leading to bilayer density imbalance), both promoted by asymmetric protonation changes of lipid headgroups. Whereas the occurrence of the first mechanism only requires the presence of a protonable charged phospholipid, the second mechanism requires the occurrence of an uncharged form of this phospholipid, as well as a moderately sized headgroup. Apart from PG, phosphatidic acid~\cite{Eastman1991} and phosphatidylinositol are likely candidates. Both mechanisms lead to different relaxation properties, that influence the dynamics of shape changes. Furthermore, the occurrence of a flip-flop mechanism for PG may explain the particular temperature-sensitivity of mitochondrial function in the cdr1-null mutant. In a previous study, it was found that PG flip-flop has an activation energy of 31 kcal/mole, corresponding to a threefold increase of flip-flop rate every 7$^{\circ}$C. This is likely to affect the mutant cristae morphology in vivo and therefore mitochondrial function. Therefore, this study gives further support to our suggestion of an important role of the lipid composition in the MIM structure and function. Additionally, this study contributes to the general understanding of membrane tubulation, which occurs in processes as diverse as Golgi remodeling~\cite{Ha2012}, endosome sorting~\cite{Breusegem2014}, filopodia~\cite{Daniels2013} and microvilli~\cite{Maheshwari2004} formation, cell-cell contacts~\cite{Okochi2009}, and intercellular communications~\cite{Gerdes2008}.

\subsection{Alzheimer's disease and the failure to form mitochondrial cristae}
\label{Alzh}
Alzheimer's disease (AD) is a degenerative disease of the central nervous system that causes massive neuronal loss, leading inevitably to progressive degeneration of mental capacities, and, ultimately, to death~\cite{Mattson2004}. It is recognized that the progress of AD leads to morphological and functional neuronal pathologies at the extracellular as well at the intracellular level, both levels being non-exclusive and probably coupled~\cite{LaFerla2007, Wirths2004}. It is clear now that the increased level of amyloid-$\beta$ (A$\beta$) peptide monomers, mainly A$\beta$ (1-42), although not inherently harmful, leads to the formation of a diversity of A$\beta$ aggregates (oligomers, proto-fibrils, fibrils) which are involved in a variety of pathological mechanisms when accumulating in the extra- and/or intracellular space of neurons. Recently, a surge in studies on AD-related neuronal pathologies focused on A$\beta$-induced mitochondrial dysfunctions and morphological alterations~\cite{Swerdlow2010, Swerdlow2004, Castellani2002, Manczak2006, Reddy2008, Caspersen2005, Chen2010}. Our work~\cite{Khalifat2012} examined these problems using, for the first time, a bio-mimetic artificial membrane approach. Furthermore, it was shown that A$\beta$ can be imported from the cytoplasm into mitochondria via the translocase of the outer mitochondrial membrane machinery~\cite{Petersen2008}. The same study also demonstrated that A$\beta$ accumulates in the mitochondrial cristae both in vivo and in vitro. However, with the recent localization of the A$\beta$ precursor protein (A$\beta$PP) in mitochondrial membrane fraction from brain of AD cases~\cite{Yamaguchi1992}, the possibility of in situ A$\beta$ production within mitochondria should be explored. In comparison with normal specimens, mitochondria from AD patients show a larger variety of shapes and sizes as well as altered cristae morphology. A number of electron microscopy morphometric studies have shown mitochondria with broken (disrupted) cristae~\cite{Hirai3017, Baloyannis2006, doi:10.1177/153331750401900205}. Surprisingly, the mechanisms relating the accumulation of A$\beta$ in mitochondrial cristae and matrix to the large number of mitochondria with broken and sparse cristae observed in AD patients' neurons were not evoked. The problem is a challenging one, as it is now widely recognized that mitochondrial functions determine the IM morphology and, conversely, that IM morphology can influence mitochondrial functions~\cite{Lea1994, Mannella1994, Mannella2006, Logan2006, Mannella2001, Mannella2006a}. What could be the factors that determine the failure of mitochondrial IM in the case of AD? Our hypothesis is the following: A$\beta$-induced alterations of the purely physical properties of the lipid bilayer could be the direct cause for this failure. We tested this hypothesis using an artificial model system as described below. In our work~\cite{Khalifat2012}, we again use GUVs as a biomimetic system in order to study the mitochondrial IM. We tested the effect of A$\beta$ (1-42) in different states of aggregation, ``fresh"-mostly monomeric A$\beta$, and ``aged"-mostly fibril A$\beta$ solutions, performing direct real time visualization of lipid membrane morphological changes induced by the peptide.

\begin{figure}[htb]
	\begin{center}
		\includegraphics[width=1\columnwidth]{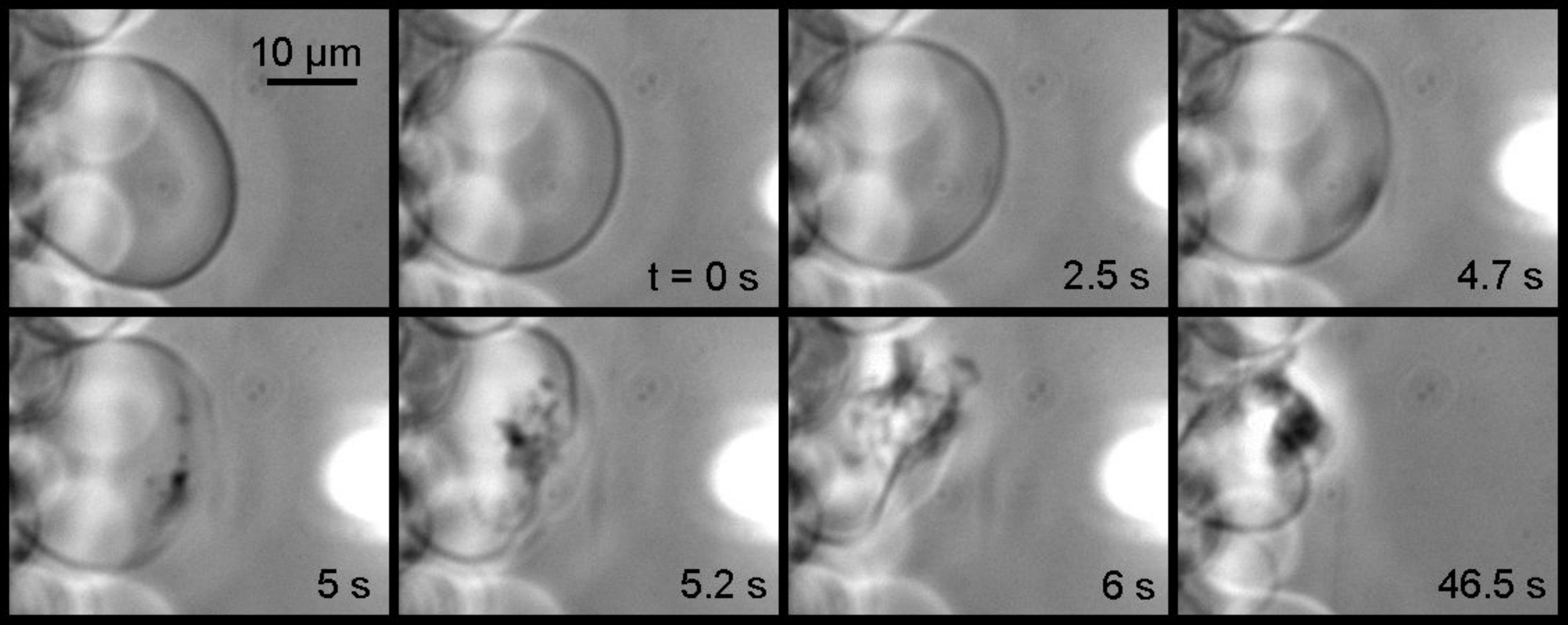}
		\caption{Test of the capacity of a GUV with molar composition EggPC/EggPE/Heart bovine CL 60:30:10 mol/mol, pre-treated with ``aged" A$\beta$(1-42) buffer solution, to develop cristae-like morphology upon local acidification. Adding locally 10 mM HCl (outside the GUV by a micropipette) initiated local membrane deformation (frame: 4.7 s), followed by harsh rupture of the membrane zone affected by the acid (frame: 5.2 s), and explosion of the GUV (frame: 6 s). As soon as acid addition was stopped, the damaged membrane zone ``re-healed", forming dense clumps (frame: 46.5 s). No cristae-like morphology developed. T=25$^{\circ}$C, pH 7.4. \textit{Reprinted from Khalifat et al.~\cite{Khalifat2012} with permission of IOS Press.}}
		\label{Ab_aged}
	\end{center}      
\end{figure}

\begin{figure}[htb]
	\begin{center}
		\includegraphics[width=0.9\columnwidth]{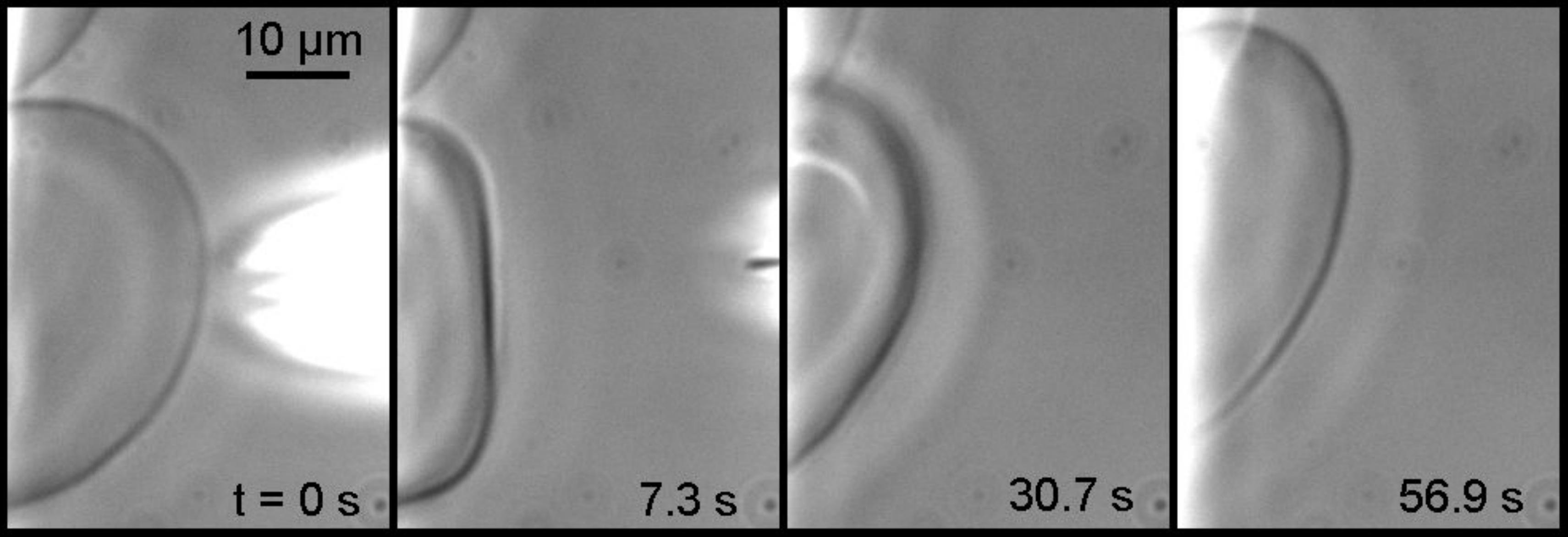}
		\caption{Test of the capacity of a GUV with molar composition EggPC/EggPE/Heart bovine CL 60:30:10 mol/mol, pre-treated with ``fresh" A$\beta$(1-42) buffer solution, to develop cristae-like morphology upon local acidification. The local acidification (with up to 100 mM HCl) induced no visible effects to the GUV membrane, in contrast with the brutal rupture (achieved adding HCl solution of only 10 mM), observed in the case of acidification, after treatment with ``aged" A$\beta$(1-42) buffer solution (Fig.~\ref{Ab_aged}). On the other hand, local acidification of GUV pre-treated with ``fresh" A$\beta$(1-42)  buffer solution did not induce any local membrane invagination and no cristae-like morphology was developed. This is in contrast with the case of local acidification of GUV not pre-treated with any A$\beta$ which does develop cristae-like morphology, as in Fig.~\ref{Nada} and Fig.~\ref{tube}, and~\cite{Khalifat2008}, all other parameters being equal. T=25$^{\circ}$C, pH 7.4. \textit{Reprinted from Khalifat et al.~\cite{Khalifat2012} with permission of IOS Press.}}
		\label{Ab_fresh}
	\end{center}      
\end{figure}

We showed that A$\beta$(1-42) renders the membrane incapable of supporting the dynamics of tubular structure formation when a local pH gradient occurs, ultimately leading to the failure of cristae-like dynamic morphology, see Figs.~\ref{Ab_aged} and~\ref{Ab_fresh} in contrast to Figs.~\ref{Nada} and~\ref{tube}. This was the case even when no problems were observed regarding membrane static behavior. In order to elucidate the molecular mechanisms underlying the failure of tubular structure formation, we carried out fluorescence studies of membrane properties involving A$\beta$(1-42)-interacting with artificial membranes mimicking the mitochondrial IM (designed as large unilamellar vesicles (LUVs)). Steady-state fluorescence emission measurements of Prodan and Laurdan generalized polarization, in relation with lipid bilayer order and hydration, as well as measurements of DPH anisotropy (rDPH), in relation with lipid bilayer fluidity, showed that A$\beta$(1-42)-fibrils induce lipid bilayer dehydration and decreasing membrane fluidity in LUVs that model mitochondrial IM. These results suggest that dramatic changes in membrane dynamic properties at the scale of the entire GUV result from A$\beta$(1-42)-induced membrane physical property changes. A$\beta$(1-42)-induced changes in dynamic friction between the two leaflets of lipid membrane may take place as well.

Our work puts forward several new ideas which might be biologically relevant. First, the A$\beta$ fibril aggregates might change directly (without lipid biochemical degradation) the global physical properties of the mitochondrial IM by decreasing membrane core fluidity, inducing lipid bilayer dehydration, and increasing apolarity (structuring) close to the water/lipid interface of the lipid bilayer. Second, we suggest that changes in lipid bilayer physical properties induced by A$\beta$ fibrils may play a special direct role in membrane functions involving rapid membrane shape changes, this being typically the case of the mitochondrial IM. In fact, the membrane, being ``normally" visco-elastic, becomes more viscous then elastic and loses its capacity to respond quickly to the mechanical constraint imposed by local acidification. The direct consequence of these alternations could be a harsh membrane rupture due to the rapid membrane shape changes driven by local acidification. 

The mechanisms affecting mitochondria in AD remain controversial: any deterioration of mitochondrial functions may cause neuronal deficiency. Our biomimetic studies involving artificial membranes encourage us to put forward an original hypothesis regarding mitochondria deficiency and AD: AD might be a consequence of the failure of fundamental physical, purely mechanical properties of the mitochondrial IM. We will call the specific A$\beta$ toxicity in this case ``mechanical toxicity''. Thus, A$\beta$-induced ``mechanical toxicity'' could lead to a membrane inability to support the shape dynamics underlying, and inherent to, normal functioning of the mitochondrial IM. In addition, A$\beta$ fibril-induced membrane dehydration, ordering, and increased membrane viscosity might cause membrane protein malfunctions, thereby contributing indirectly to the insufficiency of respiratory chain protein functions and ATP synthesis. On the other hand, A$\beta$ oligomers might induce ion channels formation, compromising the driving force behind the characteristic tubular cristae morphology, abolishing mitochondrial transmembrane potential, decreasing the capacity to accumulate calcium, and uncoupling respiration.

\section{Conclusion}
In this review, we described the influence of pH gradients and local variations on vectorial motional processes in model membrane systems. These phenomena provide an original exeample of how a molecular property, namely the ability of a specific lipid to bind or release a proton or hydroxyl ion, can have a long-range effect on the structure and dynamics of a whole molecular assembly such as a bilayer membrane. Such results from experiments and theory, developed for model systems, shed light on possible specific roles of dynamic molecular processes involving lipids in complex behavior in biomembranes underlying the build up and regulation of integrated biological functions.

\section*{Acknowledgement}
This work was supported by the Core-to-Core Program ``Non-equilibrium dynamics of soft matter and information" from the Japan Society for the Promotion of Science. We also thank the joint project ``Campus France No.38669ZB PHC-Rila-Bulgarie" (from the Bulgarian side DHTC-France 01-4) and the LABEX ``Who I Am?".

\bibliographystyle{model1-num-names}

\end{document}